\documentclass[a4paper,11pt]{article}
\usepackage{a4wide,amsmath,amssymb,bm,bbm}
\usepackage[makeroom]{cancel}
\usepackage[english]{babel}
\usepackage[utf8]{inputenc}

\usepackage{color}

\usepackage{hyperref,enumerate}
\hypersetup{
    colorlinks,
    linkcolor={black},
    citecolor={black},
    urlcolor={black}
}

\DeclareMathOperator{\tr}{tr}
\renewcommand{\o}{\overline}
\renewcommand{\u}{\underline}

\makeatletter

\@addtoreset{equation}{section} \makeatother

\begin{document}
\setcounter{page}{0}

\hfill
\vspace{30pt}

\begin{center}
{\huge{\bf {Completing $R^4$ using $O(d,d)$}}}

\vspace{80pt}

Linus Wulff

\vspace{15pt}

\small {\it Department of Theoretical Physics and Astrophysics, Faculty of Science, Masaryk University\\ 611 37 Brno, Czech Republic}
\\
\vspace{12pt}
\texttt{wulff@physics.muni.cz}\\

\vspace{80pt}

{\bf Abstract}
\end{center}
\noindent
The tree-level string effective action is known to contain quartic Riemann terms with coefficient $\zeta(3)\alpha'^3$. In the case of the type II string this is the first $\alpha'$ correction. We use the requirement that the action reduced on a $d$-torus should have an $O(d,d)$ symmetry to find the B-field couplings up to fifth order in fields. The answer turns out to have a surprisingly intricate structure.

\clearpage
\tableofcontents

\section{Introduction and summary of results}
The string theory effective action has a double expansion in the inverse string tension $\alpha'$ and the string coupling $g_s$. Here we will consider tree-level string theory and so ignore all $g_s$ corrections. The tree-level effective action has a very interesting property -- its dimensional reduction to $D-d$ dimensions ($D=10$ or 26 being the critical dimension) has a continuous $O(d,d;\mathbbm R)$ symmetry \cite{Meissner:1991zj,Meissner:1991ge}, which extends to all orders in $\alpha'$ \cite{Sen:1991zi}. Our goal here is to use this symmetry to learn about the structure of $\alpha'$ corrections. Specifically, we will focus on the first $\alpha'$ correction which is common to all string theories. The metric terms have been known for a long time and take the form \cite{Gross:1986iv,Grisaru:1986vi,Freeman:1986zh,Cai:1986sa,Gross:1986mw,Jack:1988sw}
\begin{equation}
S^{(3)}=\frac{\alpha'^3\zeta(3)}{3\cdot 2^{13}}\int d^Dx\sqrt{-G}\,e^{-2\Phi}(t_8t_8R^4+\tfrac14\varepsilon_8\varepsilon_8R^4)\,,
\label{eq:alpha3correction}
\end{equation}
where $t_8t_8R^4$ is shorthand for
\begin{equation}
t_{8\,a_1\cdots a_8}t_8^{b_1\cdots b_8}R^{a_1a_2}{}_{b_1b_2}R^{a_3a_4}{}_{b_3b_4}R^{a_5a_6}{}_{b_5b_6}R^{a_7a_8}{}_{b_7b_8}
\end{equation}
and similarly for $\varepsilon_8\varepsilon_8R^4$. These tensor structures are defined as
\begin{equation}
\varepsilon_{8\,a_1\cdots a_8}\varepsilon_8^{b_1\cdots b_8}=\tfrac12\varepsilon_{a_1\cdots a_8cd}\varepsilon^{b_1\cdots b_8cd}
\end{equation}
and
\begin{equation}
t_{abcdefgh}M_1^{ab}M_2^{cd}M_3^{ef}M_4^{gh}
=
8\tr(M_1M_2M_3M_4)
-2\tr(M_1M_2)\tr(M_3M_4)
+\mathrm{cyclic}(234)
\label{eq:t8}
\end{equation}
for anti-symmetric matrices $M_{1,2,3,4}$. It is important to note that the second term in (\ref{eq:alpha3correction}) is a total derivative at the leading order in fields and throwing away the total derivative we may write $\varepsilon_8\varepsilon_8R^4\sim\omega^2R^3$ ignoring terms of higher than fifth order in fields.

Here we will use the requirement of $O(d,d)$ symmetry of the reduced action to fix the couplings involving the $B$-field up to the fifth order in fields. We will see that $O(d,d)$ requires a surprisingly intricate form for these couplings. The full set of couplings of the NS sector fields have been previously found in \cite{Garousi:2020mqn,Garousi:2020gio} by a brute force calculation -- writing the most general ansatz in ten dimensions and requiring T-duality symmetry of the circle reduction.\footnote{A cosmological reduction of all spatial dimensions has also been considered \cite{Codina:2020kvj,Codina:2021cxh,David:2021jqn}, but this is not enough to fix the form of the $D$-dimensional action.} This was shown to lead to a unique result. Unfortunately, the resulting action is extremely complicated and it is very hard to see any structure in it. This is the reason we revisit the calculation here using tools adapted to the $O(d,d)$ symmetry and finding a simpler, though still complicated, form for the effective action. We find the following form for the effective action (up to the overall coefficient)
\begin{equation}
L=L_{\hat R^4}+L_{(\omega^2+H^2)R^3}+L_{(H\wedge H)R^3}+L_{H^2\nabla H^2R}+\ldots\,,
\end{equation}
where the ellipsis denotes terms involving the dilaton and RR-fields, which we don't determine, and terms of sixth and higher order in fields. These couplings have the following form. First we have
\begin{equation}
L_{\hat R^4}=\frac{1}{16}t_8t_8\hat R^4
\end{equation}
where we have defined
\begin{equation}
\hat R^{ab}{}_{cd}=R^{(-)ab}{}_{cd}-\tfrac12H^{abe}H_{ecd}=R^{ab}{}_{cd}
-\nabla^{[a}H^{b]}{}_{cd}
+\tfrac12H^{[a}{}_{ce}H^{b]e}{}_{d}
-\tfrac12H^{abe}H_{ecd}
\end{equation}
and $R^{(\pm)}$ is the Riemann tensor computed from the torsionful spin connection $\omega_a^{(\pm)bc}=\omega_a{}^{bc}\pm\frac12H_a{}^{bc}$. The second contribution is given by
\begin{equation}
\begin{aligned}
L&_{(\omega^2+H^2)R^3}
=
-\frac1{64}\varepsilon_9\varepsilon_9[\omega^2+H^2]\hat R^3
\\
=&
\frac1{64}\varepsilon_8\varepsilon_8(R^{(-)})^4
+\frac1{64}\varepsilon_{a_1\cdots a_9}\varepsilon^{b_1\cdots b_9}\left(\tfrac{5}{9}H^{a_1a_2a_3}H_{b_1b_2b_3}-H_{b_1}{}^{a_1a_2}H^{a_3}{}_{b_2b_3}\right)\left(\hat R^3\right)^{a_4\cdots a_9}{}_{b_4\cdots b_9}
+\ldots
\end{aligned}
\label{eq:eeR4}
\end{equation}
where the term in brackets in the first line stands for
\begin{equation}
\tfrac{1}{5}\left(\omega_{b_1}^{(+)a_1a_2}\omega^{(-)a_3}{}_{b_2b_3}-[\omega^{(+)}-\tfrac13H]^{a_1a_2a_3}[\omega^{(-)}+\tfrac13H]_{b_1b_2b_3}\right)
+H_{b_1}{}^{a_1a_2}H^{a_3}{}_{b_2b_3}-\tfrac{1}{9}H^{a_1a_2a_3}H_{b_1b_2b_3}
\end{equation}
and the ellipsis denotes total derivatives and terms of sixth order or higher in fields. In particular, we reproduce the $\varepsilon_8\varepsilon_8R^4$ term with the correct coefficient.
The equality of the two expressions is shown in appendix \ref{app:eeR4}. Then we have additional $H^2R^3$-terms which take the form
\begin{align}
L_{(H\wedge H)R^3}
=&\,
\frac{6!^2}{6^38}H_{abc}H^{def}\hat R^{[ag}{}_{[dh}\hat R^{bh}{}_{ek}\hat R^{ck]}{}_{fg]}
+\frac{5!}{2}H_{abc}H_{de}{}^f\hat R^{[ad}{}_{gh}\hat R^{be|hk|}\hat R^{c]g}{}_{fk}
\nonumber\\
&{}
+\frac{5!}{2}H^{abc}H^{de}{}_f\hat R^{gh}{}_{[ad}\hat R_{|hk|be}\hat R^{fk}{}_{c]g}\,.
\label{eq:Lprime}
%
%
%
%
\end{align}
Note that they do not contain any contractions between the $H$'s. The need for such terms was seen from amplitude calculations in \cite{Liu:2019ses}. Finally, we have the terms of the form $H^2\nabla H^2R$, which are by far the most complicated. They take the form
\begin{equation}
L_{H^2\nabla H^2R}=
6H^{abc}H^{def}\nabla^kH_{cde}\nabla_aH_{bgh}R^{gh}{}_{kf}
+3H^{abc}H^{def}\nabla^kH_{cde}\nabla_kH_{fgh}R^{gh}{}_{ab}
+\frac{3\cdot4!}{2}(L_1+L_2)\,,
\label{eq:Lbis}
\end{equation}
where $L_1$ and $L_2$ are distinguished by the structure of the contractions and are given in (\ref{eq:L1}) and (\ref{eq:L2}). The total number of these terms is 42 and their structure is surprisingly intricate. Still, compared to the 106 terms of this form in \cite{Garousi:2020gio}, we have clearly achieved some simplification.\footnote{Curiously, while the complicated $(H\wedge H)R^3$ and $H^2\nabla H^2R$ terms found above are required at tree-level by $O(d,d)$, they are absent at one loop \cite{Richards:2008sa}. The one-loop $R^4$-terms therefore seem to have a much simpler structure than the tree-level ones, even though in the type IIB case the purely metric terms are exactly the same. In particular this means that there must be several supersymmetric $R^4$ invariants, as already argued in \cite{Liu:2019ses}.}
 
Ignoring terms of order $H^4$ it is easy to see that our results match precisely those of \cite{Liu:2019ses},\footnote{Except that our $L_{(H\wedge H)R^3}$ is 8 times that of \cite{Liu:2019ses}.} which determined all the $H^2$ couplings using string amplitude calculations.\footnote{Note that we may replace $H^{a_1a_2a_3}H_{b_1b_2b_3}\rightarrow 3H_{b_1}{}^{a_1a_2}H^{a_3}{}_{b_2b_3}$ in (\ref{eq:eeR4}) up to $H^4$ terms, as follows from a similar calculation to (\ref{eq:omega-H-id}) with $\omega^{(+)}$ replaced by $H$.} Due to the very complicated form of the $H^4$ terms in \cite{Garousi:2020gio}, we have not attempted a comparison of these.

To derive this result our strategy is the following. We start with the known $t_8t_8R^4$ term in $D=10$ (or $D=26$). Then we use ideas from Double Field Theory (DFT) \cite{Siegel:1993th,Hull:2009mi,Hohm:2010pp} to rewrite it in terms of an $O(D,D)$ invariant analog of the Riemann tensor. This object is not Lorentz invariant and we have to add terms quadratic in the spin connection to compensate for this. These extra terms can also be expressed in terms of objects from DFT. In doing so we obtain an expression which looks $O(D,D)$ invariant, except for the fact that the double Lorentz symmetry needed to have a consistent DFT formulation is explicitly broken. Only its diagonal, the usual Lorentz group, is preserved. It is important to note that we are working only with completely gauge-fixed objects from DFT, which can always be expressed only in terms of the usual metric/vielbein and $B$-field. Therefore, the explicit breaking of the DFT symmetries does not lead to any inconsistencies. It seems that we could just as well work with the usual metric and vielbein, rather than involve the DFT notation. However, the reason for using the DFT notation is 1) that the dimensional reduction of the action expressed in terms of the DFT fields to $D-d$ dimensions is simple to perform, but more importantly 2) that one can read off directly which terms in the reduced action are compatible with $O(d,d)$ and which terms are not. More precisely, we work with a frame-like formulation where the global $O(d,d)$ symmetry is made manifest at the cost of introducing a local (internal) double Lorentz symmetry $O(d)\times O(d)$ which is not manifest, but needed for consistency. We require that the terms in the reduced action which would explicitly violate the $O(d)\times O(d)$ symmetry, by having an index transforming under the first factor contracted with an index transforming under the second factor, should cancel. {This is a very strong requirement and, in fact, we argue that at least in the present case it is equivalent to $O(d,d)$ invariance. We find} that the required cancellations are only possible if one adds particular terms involving the NSNS field strength $H$ to the $D$-dimensional action. We determine these by working order by order in $H$. To simplify the calculations we make the following assumptions
\begin{itemize}
	\item[1.] We look only at the terms in the reduced action quadratic in the gauge vectors and not containing the internal scalars.
	\item[2.] We ignore terms involving the dilaton.
	\item[3.] We use the equations of motion in the reduced theory, i.e. we allow field redefinitions after reduction.
\end{itemize}
Regarding the first point, it is not hard to see that the remaining terms, i.e. terms quartic in the gauge vectors or terms containing scalars, will cancel along very similar lines, though these are typically less constraining. The second assumption means that we cannot determine any of the couplings involving the dilaton. With some extra work one can of course go back and determine them by keeping track of them everywhere. Finally, regarding the last point, ideally one would like to allow only field redefinitions in the $D$-dimensional theory, but we did not investigate this as the calculations become more complicated. We also did not attempt to prove that the result is unique (up to field redefinitions), since this already follows from \cite{Garousi:2020gio}.

Let us emphasize again that, while some of our expressions are written using a mix of DFT and standard notation, this is just a trick to simplify the bookkeeping and we are always working with the standard gravity fields and symmetries. In any expression where the generalized fluxes $F$ appear they are understood to be expressed in terms of the usual spin connection and $H$ as in (\ref{eq:Fs}), i.e. the DFT symmetries are completely gauge-fixed. {However, from our results it is straightforward to extract a non gauge-fixed DFT description of the \emph{reduced} theory, where only the \emph{internal} coordinates are doubled. All one needs to do is keep all the $O(d,d)$ compatible terms in the dimensional reduction and forget about the DFT gauge fixing of the internal coordinates. One should also include the scalars which we set to zero. We did not try to write the resulting action since it would contain quite a large number of terms and our main interest here is the original $D$-dimensional action.}

It might seem that we could have worked instead within DFT from the beginning, but we believe this is actually not possible. Indeed, in \cite{Hronek:2020xxi} it was shown that while the $R^4$-terms can be cast in $O(D,D)$ invariant DFT form at the quartic order in fields, it is not possible to complete them (within DFT and with some mild assumptions) by terms of fifth order in fields. This might seem surprising given the fact that the lower order $\alpha'$ and $\alpha'^2$ corrections to the bosonic and heterotic string \emph{can} be cast in DFT form \cite{Marques:2015vua,Hronek:2021nqk} (see also \cite{Hohm:2013jaa,Hohm:2014xsa} for earlier attempts). However, the reason is that these lower corrections (together with an infinite tower of higher corrections) can be generated from an \emph{uncorrected} extended gauged DFT action, by imposing an identification of the gauge field and spin connection \cite{Baron:2018lve,Baron:2020xel} (see also \cite{Lee:2015kba}), a la Bergshoeff and de Roo \cite{Bergshoeff:1988nn,Bergshoeff:1989de}. There is no similar trick for generating the $\zeta(3)\alpha'^3$ corrections we are interested in here. Indeed, our calculations show explicitly how terms that are not compatible with an $O(D,D)$ invariant DFT description in $D$ dimensions can lead, upon dimensional reduction to $D-d$ dimensions, to terms which \emph{are} compatible with and $O(d,d)$ invariant DFT description of the reduced theory, thanks to additional cancellations possible only after dimensional reduction. Note that the difference between the $O(D,D)$ and $O(d,d)$ invariant case is not just that $d<D$, the more important difference is that in the latter case there are $d$ isometries, which are `rotated' by $O(d,d)$, while in the former case no isometries are assumed, which is much more restrictive.

The remainder of the paper is organized as follows. In section \ref{sec:DFT} we introduce the DFT parametrization of the fields that we will use. Then in section \ref{sec:reduction} we discuss the dimensional reduction in terms of these fields. The main part of the paper is section \ref{sec:Odd} where we require the non-invariant terms in the reduced action to cancel, fixing the form of the $D$-dimensional action. We end with some conclusions. Details of the calculations are provided in the appendix.

\section{DFT parametrization of fields}\label{sec:DFT}
Here we introduce the necessary concepts from DFT. As the name suggests DFT involves doubling the spacetime coordinates $x\rightarrow (\tilde x,x)$. One then imposes an $O(D,D)$ invariant ``section condition'' which effectively removes half of them. Here we will mostly ignore the doubling and work with the solution to the section condition where the additional coordinates $\tilde x$ are set to zero. In fact, in the rest of the paper we will work only with completely gauge-fixed DFT, which is equivalent to the usual gravity description. The reason for still using DFT notation is that it provides a natural way to organize the fields in order to recognize directly which terms in the reduced action are compatible with $O(d,d)$ symmetry and which are not.

We will use the so-called flux formulation of \cite{Geissbuhler:2013uka}, building on the frame-like formulation of DFT \cite{Hohm:2010xe}. The basic field is the generalized vielbein
\begin{equation}
E_A{}^M=
\frac{1}{\sqrt2}
\left(
\begin{array}{cc}
e^{(+)a}{}_m-e^{(+)an}B_{nm} & e^{(+)am}\\
-e^{(-)}_{am}-e^{(-)}_a{}^nB_{nm} & e^{(-)}_a{}^m
\end{array}
\right)\,.
\label{eq:E}
\end{equation}
It is constructed from two sets of vielbeins $e^{(\pm)}$ for the metric $G_{mn}$, which transform independently as $\Lambda^{(\pm)}e^{(\pm)}$ under two copies of the Lorentz group, and the $B$-field.\footnote{
The dilaton $\Phi$, which will not play any role here, is encoded in the generalized dilaton $d$ defined as
$$
e^{-2d}=e^{-2\Phi}\sqrt{-G}\,.
$$
} The standard supergravity fields are recovered by fixing the gauge $e^{(+)}=e^{(-)}=e$, leaving only the diagonal copy of the Lorentz group. In this formulation a global $O(D,D)$ symmetry acting on the doubled coordinate index $M=({}^m,{}_m)$ is manifest. Instead, consistency requires the local double Lorentz symmetry $O(D-1,1)\times O(D-1,1)$ acting on the index $A=({}^a,{}_a)$, which is not manifest, to be preserved.

There are two constant metrics, the $O(D,D)$ metric $\eta^{AB}$ and the generalized metric $\mathcal H^{AB}$, which take the form
\begin{equation}
\eta^{AB}=
\left(
\begin{array}{cc}
	\eta_{ab} & 0\\
	0 & -\eta^{ab}
\end{array}
\right)\,,\qquad
\mathcal H^{AB}=
\left(
\begin{array}{cc}
	\eta_{ab} & 0\\
	0 & \eta^{ab}
\end{array}
\right)\,,
\label{eq:eta}
\end{equation}
where $\eta=(-1,1,\ldots,1)$ is the $D$-dimensional Minkowski metric. The $O(D,D)$ metric is used to raise and lower indices. The projection operators
\begin{equation}
P_\pm^{AB}=\frac12\left(\eta^{AB}\pm\mathcal H^{AB}\right)\,,
\label{eq:Ppm}
\end{equation}
are easily seen to project on upper and lower indices respectively. The analog of the spin connection is the ``generalized flux''\footnote{Here we have defined $\partial_A=E_A{}^M\partial_M$ where $\partial_M=(0,\partial_m)$ after solving the section condition in the standard way.}
\begin{equation}
F_{ABC}=3\partial_{[A}E_B{}^ME_{C]M}\,.
\label{eq:fluxes}
\end{equation}
Since we can use the projection operators to project onto upper or lower indices we actually have four objects. After fixing the double Lorentz symmetry by imposing the gauge $e^{(+)}=e^{(-)}$ (and solving the section condition) they reduce to
\begin{align}
&F^a{}_{bc}=\frac{1}{\sqrt2}\omega^{(-)a}{}_{bc}\,,\qquad 
F_a{}^{bc}=-\frac{1}{\sqrt2}\omega_a^{(+)bc}\,,
\nonumber\\
&F_{abc}=\frac{1}{\sqrt2}(3\omega^{(-)}_{[abc]}+H_{abc})\,,\qquad 
F^{abc}=-\frac{1}{\sqrt2}(3\omega^{(+)[abc]}-H^{abc})\,.
\label{eq:Fs}
\end{align}
By construction $F_{ABC}$ is invariant under constant $O(D,D)$ transformations since these simply rotate the coordinate indices $M,N,\ldots$. However, it transforms similarly to a connection under the $O(D-1,1)\times O(D-1,1)$ double Lorentz transformations acting on the indices $A,B,\ldots$. In particular, after splitting the indices into upper and lower ones using the projectors, the upper indices are rotated by the first Lorentz group while the lower indices are rotated by the second. This means that $F$'s with different index placements, e.g. $F_a{}^{bc}$ and $F^{abc}$, are independent fields (in DFT) since they transform differently.\footnote{After fixing the gauge $e^{(+)}=e^{(-)}$ they are no longer independent, as is clear from (\ref{eq:Fs}).} We are therefore not allowed to raise and lower the indices on these fields. Importantly for our later discussion it also means that a contraction of two indices with $\eta_{ab}$, e.g.
\begin{equation}
\eta^{ad}F_a{}^{bc}F_{def}\,,
\end{equation}
is compatible with the double Lorentz symmetry, since the contracted indices transform under the same group. On the other hand a contraction of an upper and a lower index, e.g.
\begin{equation}
F_a{}^{bc}F^{ade}\,,
\end{equation}
would explicitly break the symmetry, since the two contracted indices transform under \emph{different} Lorentz groups. It is terms of this form (with the contracted index an internal index) that we will require to cancel in the reduced theory.

We can also introduce a DFT analog of the Riemann tensor. Following \cite{Hronek:2020skb} we define\footnote{Defining $\mathcal R_{ab}{}^{cd}$ similarly we have $\mathcal R_{ab}{}^{cd}=-\mathcal R^{cd}{}_{ab}$.}
\begin{equation}
\mathcal R^{ab}{}_{cd}=2\partial^{[a}F^{b]}{}_{cd}-\eta_{ef}F^{abe}F^f{}_{cd}+2\eta^{ef}F^{[a}{}_{ce}F^{b]}{}_{fd}\,.
\label{eq:R}
\end{equation}
When we fix the gauge $e^{(+)}=e^{(-)}$ this reduces to
\begin{equation}
\mathcal R^{ab}{}_{cd}
=
\frac12
(
R^{(-)ab}{}_{cd}
+\omega^{(+)eab}\omega^{(-)}{}_{ecd}
)\,,
\end{equation}
which shows that unlike the usual Riemann tensor this object is \emph{not} Lorentz covariant. Conversely, we can instead take the combination
\begin{equation}
\mathcal R^{ab}{}_{cd}+F_e{}^{ab}F^e{}_{cd}\,,
\label{eq:R'}
\end{equation}
which is Lorentz covariant, in fact it reduces to $\frac12R^{(-)ab}{}_{cd}$ on setting $e^{(+)}=e^{(-)}$, but is not compatible with double Lorentz symmetry due to the contraction of an upper and a lower index in the second term. It therefore only makes sense to work with this object after gauge-fixing the DFT symmetries.

\section{Dimensional reduction}\label{sec:reduction}
We will denote $D$-dimensional quantities by calligraphic letters in order to distinguish them from the corresponding quantities in the dimensionally reduced theory. We take the following dimensional reduction ansatz for the generalized vielbein
\begin{equation}
\mathcal{E_A{}^M}
=
E_{\mathcal A}{}^{\mathcal N}(1+U)_{\mathcal N}{}^{\mathcal M}\,,
\end{equation}
where $E$ is diagonal with non-zero components 
\begin{equation}
E_A{}^B\qquad\mbox{and}\qquad E_{A'}{}^{B'}
\end{equation}
while the non-zero components of $U$ are
\begin{equation}
U_{M'n}=A_{M'n}\,,\qquad
U_m{}^{N'}=-A^{N'}_m\,,\qquad
U_{mn}=-\tfrac12A_m^{K'}A_{K'n}\,.
\end{equation}
Note that this form guarantees that $1+U\in O(D-d,D-d)\times O(d,d)$. Internal indices are denoted with primes and the various indices and groups under which they transform are summarized in table \ref{tab:indices}. We are interested only in the internal symmetries ($O(d,d)$ and $O(d)\times O(d)$) and we will take the external part to be gauge-fixed, removing $O(D-d,D-d)$ and breaking $O(D-d-1,1)\times O(D-d-1,1)\rightarrow O(D-d-1,1)$, the usual Lorentz group for the external directions.
\begin{table}%
\begin{center}
	\begin{tabular}{lcr}
Index & Internal/External & Transforms under\\
\hline
$\mathcal M=(M,M')$ & - & Global $O(D,D)$\\
$\mathcal A=(A,A')$ & - & Local $O(D-1,1)\times O(D-1,1)$\\
$M=({}^m,{}_m)$ & External & Global $O(D-d,D-d)$\\
$A=({}^a,{}_a)$ & External & Local $O(D-d-1,1)\times O(D-d-1,1)$\\
$M'=({}^{m'},{}_{m'})$ & Internal & Global $O(d,d)$\\
$A'=({}^{a'},{}_{a'})$ & Internal & Local $O(d)\times O(d)$\\
\end{tabular}
\end{center}
\caption{Summary of index notation.}
\label{tab:indices}
\end{table}
We have grouped the two gauge fields, coming from the metric and $B$-field respectively, into an $O(d,d)$ vector
\begin{equation}
A_{mN'}=
\left(
\begin{array}{c}
A_m^{(1)n'}\\
A_{mn'}^{(2)}
\end{array}
\right)\,.
\end{equation}
Gauge fixing $e^{(+)}=e^{(-)}$ one recovers the usual Kaluza-Klein reduction ansatz.\footnote{Namely
$$
\bm e=
\left(
\begin{array}{cc}
e_m{}^a & A^{(1)n'}_{m}e_{n'}{}^{a'}\\
0 & e_{m'}{}^{a'}
\end{array}
\right)\,,\qquad
\mathcal B=
\left(
\begin{array}{cc}
B_{mn}-A_{[m}^{(1)k'}A^{(2)}_{n]k'}+A_m^{(1)k'}A_n^{(1)l'}B_{k'l'} & A^{(2)}_{mn'}+A_m^{(1)k'}B_{k'n'}\\
-A^{(2)}_{m'n}+B_{m'k'}A_n^{(1)k'} & B_{m'n'}
\end{array}
\right)\,.
%
%
%
%
%
$$
}

The dimensional reduction of the generalized flux $\mathcal{ F_{ABC}}$ becomes
\begin{equation}
\mathcal F_{ABC}=F_{ABC}+\frac32F_{[AB}^{D'}A_{C]D'}
\,,\qquad
\mathcal F_{A'BC}=-F_{A'BC}
\,,\qquad
\mathcal F_{A'B'C}=\partial_CE_{A'}{}^{M'}E_{B'M'}\,,
\label{eq:F-red}
\end{equation}
while $\mathcal F_{A'B'C'}$ vanishes. Here we have introduced the field strength of the doubled gauge field
\begin{equation}
F_{mn}^{K'}=2\partial_{[m}A_{n]}^{K'}
\end{equation}
and used the generalized vielbein to convert the indices, i.e. $A_{A'B}=E_{A'}{}^{N'}E_B{}^mA_{mN'}$ and $F_{A'BC}=E_{A'}{}^{K'}E_{B}{}^mE_{C}{}^nF_{mnK'}$. Note that this means in particular that here $A_{a'b}=\frac{1}{\sqrt2}e_b{}^mA_{mb'}$, rather than the standard definition without the $\sqrt2$. In DFT this reduction breaks the $O(D,D)$ symmetry and double Lorentz symmetry down to their internal parts, i.e. $O(d,d)$ and double Lorentz transformations (rotations) acting on the primed indices $O(d)\times O(d)$.

For the remainder of this paper we will set the scalars that arise on dimensional reduction to zero, since this will be enough for our purposes. This amounts to $E_{A'}{}^{M'}$ being constant. Since we are also ignoring the dilaton we are starting from an action in $D$ dimensions which can be expressed in terms of $H_{abc}$ and $R^{(-)ab}{}_{cd}$ and their covariant derivatives.\footnote{Note that we use $a,b,\ldots$ both for $D$-dimensional indices and for external $(D-d)$-dimensional indices. Since these never occur together in the same expression it is hopefully clear from the context which one we mean.} Our strategy is to first write this in terms of gauge-fixed DFT fields. In particular we have from (\ref{eq:Fs})
\begin{equation}
H^{abc}=\sqrt2(F^{abc}-3\eta^{d[a}F_d{}^{bc]})\,,\qquad
H_{abc}=\sqrt2(F_{abc}-3\eta_{d[a}F^d{}_{bc]})\,.
\label{eq:H}
\end{equation}
This is of course not the only way to express $H$ in terms of the $F$'s but it is the way that violates the would-be double Lorentz symmetry the least, since it involves only one $\eta$ (remember that upper and lower indices on $F$ are rotated by different groups in DFT). Similarly we may express $R^{(-)ab}{}_{cd}$ through the combination (\ref{eq:R'}) as
\begin{equation}
R^{(-)ab}{}_{cd}=2\mathcal R^{ab}{}_{cd}+2F_e{}^{ab}F^e{}_{cd}\,.
\label{eq:R-}
\end{equation}
However, it will be more convenient for our purposes to include some quadratic terms in $H$ and work instead with
\begin{equation}
\hat R^{ab}{}_{cd}=
R^{(-)ab}{}_{cd}
+aH^{abe}H_{ecd}
+bH^a{}_{e[c}H_{d]}{}^{be}\,,
\label{eq:Rhat}
\end{equation}
where $a$ and $b$ are constants to be fixed. Let us now compute the dimensional reduction of this object. We first promote it to an expression in terms of (gauge-fixed) generalized fluxes using (\ref{eq:H}) and (\ref{eq:R-}). Next, we use the reduction of the generalized flux (\ref{eq:F-red}), recalling the definition of $\mathcal R^{ab}{}_{cd}$ in (\ref{eq:R}) and letting $E_{A'}{}^{M'}$ be constant. Denoting again the $D$-dimensional $\hat R$ as $\hat{\mathcal R}$ one finds the reduction
\begin{equation}
\begin{aligned}
\hat{\mathcal R}^{ab}{}_{cd}=&\,\hat R^{ab}{}_{cd}+\Delta^{ab}{}_{cd}\,,\\
\hat{\mathcal R}^{a'b}{}_{cd}=&\,\hat R^{a'b}{}_{cd}+\Delta^{a'b}{}_{cd}\,,\\
\hat{\mathcal R}^{ab}{}_{c'd}=&\,\hat R^{ab}{}_{c'd}+\Delta^{ab}{}_{c'd}\,,\\
\hat{\mathcal R}^{a'b'}{}_{cd}=&\,\hat R^{a'b'}{}_{cd}+\Delta^{a'b'}{}_{cd}\,,\\
\hat{\mathcal R}^{ab}{}_{c'd'}=&\,\hat R^{ab}{}_{c'd'}+\Delta^{ab}{}_{c'd'}\,,\\
\hat{\mathcal R}^{a'b}{}_{c'd}=&\,\hat R^{a'b}{}_{c'd}+\Delta^{a'b}{}_{c'd}\,,
\end{aligned}
\end{equation}
while the components with more than two primed indices vanish. As discussed at the end of the previous section, the object we started with does not respect the $D$-dimensional double Lorentz symmetry. After the reduction we are interested only in the internal double Lorentz symmetry rotating the primed indices. We have therefore split the RHS into terms which are compatible with this symmetry (primed indices contracted only with $\eta^{a'b'}$) and terms which would explicitly violate it (primed indices contracted with $\delta^{a'}_{b'}$). The ones that are compatible with such a symmetry are
\begin{equation}
\begin{aligned}
\hat R^{ab}{}_{cd}
=&\,
R^{(-)ab}{}_{cd}
+aH'^{abe}H'_{ecd}
+bH'_e{}^a{}_{[c}H'^{eb}{}_{d]}
+2(a-1)\eta_{e'f'}F^{e'ab}F^{f'}_{cd}
\\
&\,{}
+2a\eta^{e'f'}F_{e'}{}^{ab}F_{f'cd}
+2b\eta_{e'f'}F^{e'a}{}_{[c}F^{f'b}{}_{d]}
+2(b-2)\eta^{e'f'}F_{e'c}{}^{[a}F_{f'd}{}^{b]}\,,
\\
\hat R^{a'b}{}_{cd}
=&\,
-\sqrt2
\left(
\nabla^{(-)b}F^{a'}{}_{cd}
+aF^{a'be}H'_{ecd}
-bF^{a'}_{e[c}H'^{eb}{}_{d]}
\right)\,,
\\
\hat R^{ab}{}_{c'd}
=&\,
\sqrt2
\left(
\nabla^{(+)}_dF_{c'}{}^{ab}
-aH'^{abe}F_{c'de}
+bF_{c'}^{e[a}H'^{b]}{}_{ed}
\right)\,,
\\
\hat R^{a'b'}{}_{cd}
=&\,
2(2-b)F^{[a'}_{ce}F^{b']e}{}_d\,,
\\
\hat R^{ab}{}_{c'd'}
=&\,
2(2-b)F_{[c'}^{ae}F_{d']e}{}^b\,,
\\
\hat R^{a'b}{}_{c'd}
=&\,
(2-b)F_{c'}^{be}F^{a'}_{ed}
-2aF^{a'be}F_{c'ed}\,,
\end{aligned}
\end{equation}
where $\nabla^{(\pm)}$ uses the spin connection $\omega_a^{(\pm)bc}$ and $H$ is everywhere replaced by 
\begin{equation}
H'_{abc}=H_{abc}-3\sqrt2F_{[ab}^{D'}A_{c]D'}\,.
\end{equation}
The precise form of these terms will not be important for us, only that they are compatible with an internal double Lorentz symmetry and so could arise from a DFT description. What will be important here is the form of the terms which would explicitly violate $O(d)\times O(d)$, by containing contractions of an upper and a lower internal index (or a raising/lowering of a free internal index by $\eta^{a'b'}/\eta_{a'b'}$). They are
\begin{equation}
\begin{aligned}
\Delta^{ab}{}_{cd}=&\,
2aF^{e'ab}F_{e'cd}
+2(1+a)F_{e'}^{ab}F^{e'}_{cd}
+4bF^{e'[a}{}_{[c}F_{|e'|}{}^{b]}{}_{d]}\,,
\\
\Delta^{a'b}{}_{cd}
=&\,
-a\sqrt2\eta^{a'f'}F_{f'}^{be}H'_{ecd}
+b\sqrt2\eta^{a'f'}F_{f'e[c}H'^{eb}{}_{d]}\,,
\\
\Delta^{ab}{}_{c'd}
=&\,
-a\sqrt2\eta_{c'f'}F^{f'}_{de}H'^{abe}
+b\sqrt2\eta_{c'f'}F^{f'e[a}H'^{b]}{}_{ed}\,,
\\
\Delta^{a'b'}{}_{cd}
=&\,
-4b\eta^{f'[a'}F_{f'e[c}F^{b']}{}_{d]}{}^e
+2b\eta^{f'[a'}F_{f'ce}\eta^{b']g'}F_{g'd}{}^e\,,
\\
\Delta^{ab}{}_{c'd'}
=&\,
-4bF_{[c'}^{e[a}\eta_{d']f'}F^{|f'|b]}{}_e
+2b\eta_{f'[c'}F^{f'ae}\eta_{d']g'}F^{g'b}{}_e\,,
\\
\Delta^{a'b}{}_{c'd}
=&\,
-2a\eta^{a'f'}F_{f'}{}^{be}F_{c'ed}
-b\eta_{c'f'}F^{f'be}F^{a'}_{ed}
-2aF^{a'be}\eta_{c'f'}F^{f'}{}_{ed}
-bF_{c'}^{be}\eta^{a'f'}F_{f'ed}
\\
&{}
-2a\eta^{a'f'}F_{f'}{}^{be}\eta_{c'g'}F^{g'}{}_{ed}
-b\eta_{c'f'}F^{f'be}\eta^{a'g'}F_{g'ed}\,.
\end{aligned}
\label{eq:Deltas}
\end{equation}
Actually, we will only need the first three expressions, because we will confine ourselves only to terms quadratic in the KK gauge field strength. Now we are ready to turn to the question of $O(d,d)$ invariance of the reduced action.

\section{Requiring \texorpdfstring{$O(d,d)$}{O(d,d)} invariance of the reduced action}\label{sec:Odd}
We wish to fix the form of the $D$-dimensional action by requiring that the reduced theory is $O(d,d)$ invariant. Actually, rather than directly requiring $O(d,d)$ invariance, we will just require that the terms which would explicitly violate it, by not being compatible with an internal double Lorentz $O(d)\times O(d)$ symmetry and therefore cannot come from a DFT formulation, cancel out. These are precisely the terms which contain contractions of an upper and a lower internal (primed) index, since these indices would have to transform differently under the two $O(d)$ factors. 

{This is clearly a necessary condition for $O(d,d)$ invariance. In fact it is also sufficient, as we will now argue. Consider the internal double Lorentz $O(d)\times O(d)$ transformation of the reduced action (promoted to a DFT action by forgetting the gauge fixing of the internal DFT fields). From the formulas in the previous section it is clear that internal (primed) indices sit only on $F^{A'}_{cd}$, the field strength of the (doubled) KK vectors. Contractions without a derivative on $F$, $F\cdot F$ (where the dot denotes contraction of the internal index), are automatically invariant since we made sure only the invariant contractions survive. Therefore we only need to check terms with a derivative on $F$ and since, in our case, we never get more than one derivative these are of the form $\nabla F\cdot F$ and $\nabla F\cdot\nabla F$. However, we must also remember to reinstate the scalars by dropping the condition $E_{A'}{}^{M'}=$ constant that we imposed in the last section. This leads to additional terms involving $\mathcal F_{AB'C'}$ given in (\ref{eq:F-red}) and the relevant fields are $\mathcal F_{ab'c'}$ and $\mathcal F_a{}^{b'c'}$ which transform as connections under the internal double Lorentz transformations. Taking all these contributions into account the internal double Lorentz variation of the reduced action becomes (setting the scalars to zero after the variation for simplicity)
\begin{equation}
\begin{aligned}
\delta L_{\mathrm{red}}=&\,{}\nabla^a\u\lambda_{a'b'}F_{a'}^{de}F_{b'}^{fg}\u U_{adefg}
+\nabla_a\o\lambda^{a'b'}F^{a'}_{de}F^{b'}_{fg}\o U^{adefg}
\\
&{}
+\nabla^a\u\lambda_{a'b'}F_{a'}^{de}\nabla^bF_{b'}^{fg}\u V_{abdefg}
+\nabla_a\o\lambda^{a'b'}F^{a'}_{de}\nabla_bF^{b'}_{fg}\o V^{abdefg}
\\
&{}
+\nabla^a\u\lambda_{a'b'}\nabla^bF_{a'}^{de}\nabla^cF_{b'}^{fg}\u W_{abcdefg}
+\nabla_a\o\lambda^{a'b'}\nabla_bF^{a'}_{de}\nabla_cF^{b'}_{fg}\o W^{abcdefg}\,,
\end{aligned}
\end{equation}
for some functions of the fields $\u U$, $\o U$, $\u V$, $\o V$, $\u W$ and $\o W$ (here we have suppressed the $\eta^{a'b'}$ contracting the primed indices, which may not be raised or lowered, unlike the unprimed ones which are ordinary external Lorentz indices). The point is now to note that the reduced action must be invariant under (standard) Lorentz transformations, since it arises from reduction of a Lorentz invariant theory. This means that, gauge-fixing DFT to go to supergravity and setting $\o\lambda=-\u\lambda=\lambda$, the above variation must vanish. Since each term is independent however (recall that $F^{a'}$ involves the vectors coming from the metric while $F_{a'}$ involves the vectors coming from the $B$-field) this requires $\u U=\o U=\u V=\o V=\u W=\o W=0$ and it follows that the Lagrangian is actually invariant under the full internal double Lorentz symmetry.
}

The cancellation of all terms with index contractions not compatible with $O(d,d)$ turns out to be a very strong requirement, which will completely fix the form of the $D$-dimensional action. In fact, it turns out to be enough to ignore the internal scalars and to consider only the terms in the reduced action which are quadratic in the gauge field strength $F^{a'}_{ab}$. As mentioned in the introduction we will further ignore the dilaton and work only up to fifth order in fields.

We start from the following ansatz for the $D$-dimensional Lagrangian\footnote{The numerical factors are introduced for convenience. We ignore the factor $e^{-2\Phi}$ since we set the dilaton to zero here.}
\begin{equation}
L=\frac{1}{16}t_8t_8\hat R^4+\frac18\varepsilon_9\varepsilon_9(F^2+H^2)\hat R^3\,.
\end{equation}
These terms are shorthand for the following expressions
\begin{align}
t_8t_8\hat R^4
=&\,
t_{a_1\cdots a_8}t^{b_1\cdots b_8}\hat R^{a_1a_2}{}_{b_1b_2}\hat R^{a_3a_4}{}_{b_3b_4}\hat R^{a_5a_6}{}_{b_5b_6}\hat R^{a_7a_8}{}_{b_8b_8}\,,
\\
\varepsilon_9\varepsilon_9(F^2+H^2)\hat R^3
=&\,
\varepsilon_{ca_1\cdots a_9}\varepsilon^{cb_1\cdots b_9}
\Big(
\frac{c}{4}\left[F_{b_1}{}^{a_1a_2}F^{a_3}{}_{b_2b_3}-\frac19F^{a_1a_2a_3}F_{b_1b_2b_3}\right]
\label{eq:F2R3}
\\
&{}
+\frac{d}{36}H^{a_1a_2a_3}H_{b_1b_2b_3}
+\frac{e}{4}H_{b_1}{}^{a_1a_2}H^{a_3}{}_{b_2b_3}
\Big)\hat R^{a_4a_5}{}_{b_4b_5}\hat R^{a_6a_7}{}_{b_6b_7}\hat R^{a_8a_9}{}_{b_8b_9}\,,
\nonumber
\end{align}
where $t_8$ is defined in (\ref{eq:t8}) and $\hat R$ in (\ref{eq:Rhat}). Note that $\hat R$ contains two free parameters $a,b$ and above we have introduced three additional free parameters $c,d,e$. These will become fixed later. The $F$'s appearing in the above expression can be written in terms of the spin connection $\omega$ and $H$ using (\ref{eq:Fs}), but we write them this way here since then we can carry out the dimensional reduction directly. The precise combination of $F$'s with different index structure is dictated by the requirement that the action should be Lorentz invariant up to a total derivative. This means that one can add a total derivative to complete these terms to $\varepsilon_8\varepsilon_8R^4$ (see the introduction).

Dimensionally reducing this Lagrangian using the results of the previous section gives rise to terms that would explicitly violate an internal double Lorentz symmetry of the following schematic form (ignoring the scalars and terms with more powers of $F$)
\begin{align}
\frac{1}{16}t_8t_8\hat R^4\rightarrow&\,t_8t_8F^2\hat R^3+t_8FH\nabla F\hat R^2\,,
\\
\frac18\varepsilon_9\varepsilon_9(F^2+H^2)\hat R^3\rightarrow&\,
\varepsilon_8\varepsilon_8F^2\hat R^3
+\varepsilon_8\varepsilon_8(\omega F+HF)\nabla F\hat R^2
+\varepsilon_8\varepsilon_8(\omega^2+H^2)\nabla F^2\hat R\,,
\end{align}
where we have kept only terms up to fifth order in fields, so $\hat R$ can be replaced by $R^{(-)}$. The last term in the second line looks very non-Lorentz covariant. To write it more covariantly we have to integrate by parts. It is convenient to organize the calculation in powers of $H$. We start by considering the double Lorentz violating terms in the reduced theory which do not contain $H$.

\subsection{Terms of order \texorpdfstring{$H^0$}{H**0}}
Setting $H=0$ and looking at the order $F^2$ terms we have, up to total derivatives and higher order terms,
\begin{equation}
\begin{aligned}
t_8t_8F^2R^3
=&\,
\frac12\left(
aF^{a'cd}F_{a'ab}
+(1+a)F_{a'}^{cd}F^{a'}_{ab}
-2bF^{a'c}{}_aF_{a'b}{}^d
\right)
(t_8t_8R^3)^{ab}{}_{cd}\,,
\\
\varepsilon_8\varepsilon_8F^2R^3
=&\,
\frac{1}{16}\left[(c+d+3e)F_{a'}^{cd}F^{a'}_{ab}+(d+3e)F^{a'cd}F_{a'ab}\right](\varepsilon_8\varepsilon_8R^3)^{ab}{}_{cd}\,,
\\
\varepsilon_8\varepsilon_8\omega F\nabla FR^2
=&\,\frac{3c}{8}F_{a'}^{cd}F^{a'}_{ab}(\varepsilon_8\varepsilon_8R^3)^{ab}{}_{cd}\,,
\\
\varepsilon_8\varepsilon_8\omega^2\nabla F^2R
=&\,
\frac{3c}{16}F_{a'}^{cd}F^{a'}_{ab}(\varepsilon_8\varepsilon_8R^3)^{ab}{}_{cd}\,.
\end{aligned}
\end{equation}
These terms explicitly violate the would-be internal Lorentz symmetry since they contain a contraction of a lower and upper primed index. To get an $O(d,d)$ invariant reduced action we must require that they cancel. Clearly the terms with the $t_8t_8$ structure and $\varepsilon_8\varepsilon_8$ structure must cancel separately. The only way this can happen is if the combinations of $F^2$-terms are such that $(F^2)_{abcd}=-(F^2)_{cdab}$, since then they give zero due to the contraction with the $R^3$ terms which are symmetric under exchanging the pairs of indices due to the symmetry of the Riemann tensor. This in turn requires the free coefficients to satisfy\footnote{In fact, this also ensures that the order $F^4$ terms vanish since the symmetries of the Riemann tensor are the same.}
\begin{equation}
a=-\frac12,\quad b=0\quad\mbox{ and }\quad d+3e=-5c\,.
\end{equation}

Having partially fixed the free parameters we can now go back to the general $H\neq0$ case and we find (to this order $\hat R=R^{(-)}$)
\begin{equation}
t_8t_8F^2\hat R^3=\frac{1}{4}t_8t_8(F_{a'}F^{a'}-F^{a'}F_{a'})\hat R^3\,,
\label{eq:t8-1}
\end{equation}
where the index structure on the first factor is $F_{a'}^{ab}F^{a'}_{cd}-F^{a'ab}F_{a'cd}$ and
\begin{align}
t_8FH\nabla F\hat R^2
=&
12t^{b_1\cdots b_8}F_{a'}^{ae}\nabla^dF^{a'}_{b_1b_2}H_{eb_3b_4}\hat R_{acb_5b_6}\hat R^c{}_{db_7b_8}
\nonumber\\
&{}
-12t_{a_1\cdots a_8}F^{a'}_{be}\nabla^dF_{a'}^{a_1a_2}H^{ea_3a_4}\hat R^{a_5a_6bc}\hat R^{a_7a_8}{}_{cd}
\nonumber\\
&{}
-3t^{b_1\cdots b_8}F_{a'}^{ae}\nabla_aF^{a'}_{b_1b_2}H_{eb_3b_4}\hat R^{cd}{}_{b_5b_6}\hat R_{dcb_7b_8}
\nonumber\\
&{}
+3t_{a_1\cdots a_8}F^{a'}_{be}\nabla^bF_{a'}^{a_1a_2}H^{ea_3a_4}\hat R^{a_5a_6}{}_{cd}\hat R^{a_7a_8dc}\,,
\label{eq:t8-2}
\end{align}
while the $\varepsilon_8\varepsilon_8$-terms are, again dropping total derivatives and higher order terms,
\begin{align}
\varepsilon_8\varepsilon_8F^2\hat R^3
=&\,{}
-\frac{c}{16}\varepsilon_8\varepsilon_8(4F_{a'}F^{a'}+5F^{a'}F_{a'})\hat R^3\,,
\label{eq:e8-1}
\\
\varepsilon_8\varepsilon_8(\omega^2+H^2)\nabla F^2\hat R
=&\,{}
\frac{3c}{16}\varepsilon_8\varepsilon_8(F_{a'}F^{a'})\hat R^3
+\frac{3e}{2}\varepsilon_8\varepsilon_8(F_{a'}F^{a'})\nabla H\nabla H\hat R
\nonumber\\
&\,{}
-\frac{8!}{6}(d+e)\nabla_{[a_1}H^{a_1a_2a_3}\nabla^{a_4}H_{a_2a_3a_4}F_{a'}^{a_5a_6}F^{a'}_{a_5a_6}\hat R^{a_7a_8}{}_{a_7a_8]}\,,
\\
\varepsilon_8\varepsilon_8(\omega F+HF)\nabla F\hat R^2
=&\,
\frac{3c}{8}\varepsilon_8\varepsilon_8(F_{a'}F^{a'})\hat R^3
-\frac{3e}{2}\varepsilon_8\varepsilon_8(F_{a'}F^{a'})\nabla H\hat R^2
\nonumber\\
&{}
\hspace{-120pt}
+7!(d+e)
\left(
H^{a_1a_2a_3}F_{a'[a_1a_2}\nabla^{a_4}F^{a'}_{a_3a_4}
+H_{[a_1a_2a_3}F_{a'}^{a_1a_2}\nabla_{a_4}F^{a'a_3a_4}
\right)\hat R^{a_5a_6}{}_{a_5a_6}\hat R^{a_7a_8}{}_{a_7a_8]}\,.
\label{eq:e8-3}
\end{align}


\subsection{Terms of order \texorpdfstring{$H^1$}{H**1}}
Again we consider the terms in the reduced theory which would not be compatible with an internal double Lorentz symmetry, but this time the ones linear in $H$. From the $\varepsilon_8$-terms we have
\begin{align}
-\frac{15c+12e}{8}\varepsilon_8\varepsilon_8(F_{a'}F^{a'})\nabla HR^2
+2\cdot7!(d+e)H^{a_1a_2a_3}F_{a'[a_1a_2}\nabla^{a_4}F^{a'}_{a_3a_4}R^{a_5a_6}{}_{a_5a_6}R^{a_7a_8}{}_{a_7a_8]}\,.
\label{eq:e8e8H1}
\end{align}
Note that in the first term we may integrate by parts to have the derivative acting on $F$ rather than $H$. The $t_8$-terms give
\begin{align}
&{}-\frac{3}{4}t_8t_8(F_{a'}F^{a'})\nabla HR^2
+12t^{a_1\cdots a_8}F_{a'}^{ab}\nabla_dF^{a'}_{a_1a_2}H_{ba_3a_4}R_{aca_5a_6}R^{cd}{}_{a_7a_8}
\nonumber\\
&{}+3t^{a_1\cdots a_8}F_{a'}^{ab}\nabla_aF^{a'}_{a_1a_2}H_{ba_3a_4}R_{cda_5a_6}R^{cd}{}_{a_7a_8}
-(F^{a'}\leftrightarrow F_{a'})\,.
\label{eq:t8H1}
\end{align}
The first step is to rewrite the first term so that the derivative is acting on $F$ instead of $H$, since all other terms can be written in that form. To start with we have
\begin{align}
t_8t_8&(F_{a'}F^{a'})\nabla HR^2
\sim
8t^{a_1\cdots a_8}F_{a'}^{ab}F^{a'}_{a_1a_2}\nabla_bH_{ca_3a_4}R^{cd}{}_{a_5a_6}R_{daa_7a_8}
\nonumber\\
&{}
+8t^{a_1\cdots a_8}F_{a'}^{ab}F^{a'}_{a_1a_2}\nabla^cH^d{}_{a_3a_4}R_{daa_5a_6}R_{bca_7a_8}
-4t^{a_1\cdots a_8}F_{a'}^{ab}F^{a'}_{a_1a_2}\nabla_cH_{da_3a_4}R^{cd}{}_{a_5a_6}R_{aba_7a_8}
\nonumber\\
&{}
+8t^{a_1\cdots a_8}\nabla_c(F_{a'}^{ab}F^{a'}_{a_1a_2})H_{ba_3a_4}R^{cd}{}_{a_5a_6}R_{daa_7a_8}
+2t^{a_1\cdots a_8}\nabla_a(F_{a'}^{ab}F^{a'}_{a_1a_2})H_{ba_3a_4}R^{cd}{}_{a_5a_6}R_{cda_7a_8}
\nonumber\\
&{}
+8t^{a_1\cdots a_8}F_{a'}^{ab}F^{a'}_{a_1a_2}H_{ba_3a_4}\nabla_cR^{cd}{}_{a_5a_6}R_{daa_7a_8}\,,
\label{eq:H1-1}
\end{align}
where `$\sim$' means up to total derivatives and higher order terms. The last term can be removed by a field redefinition since it is proportional to the equations of motion at this order. The next step is to rewrite the first three so that the derivative is acting on $F$ rather than $H$. It is convenient to start with the terms with the fewest number of `traces' (contractions of pairs of anti-symmetrized indices) and work upwards in the number of traces. The calculations are long and some details are provided in appendix \ref{app:B}. When the dust settles one finds that the contribution of the $t_8$-terms in (\ref{eq:t8H1}) can be written as
\begin{equation}
-\frac{3}{8}\varepsilon_8\varepsilon_8(F_{a'}F^{a'})\nabla HR^2
+12F_{a'ab}H_{def}Y^{a'abdef}
-12F^{a'}_{ab}H_{def}Y_{a'}{}^{abdef}\,,
\label{eq:H1-t8-simpl}
\end{equation}
where $Y^{a'abdef}$ and $Y_{a'}{}^{abdef}$ have the structure $\nabla FR^2$ and are defined in (\ref{eq:Y}). Importantly, the $Y$-terms involve no contractions between the $F$ and $H$ sitting in front. This means that they can be canceled by adding terms quadratic in $H$, with no index contracted between the two $H$'s, to the $D$-dimensional Lagrangian without introducing additional unwanted terms in the reduced theory. One finds that the following terms do the job
\begin{align}
6\Big(&
\frac83H_{abc}H^{def}\hat R^{ag}{}_{dh}\hat R^{bh}{}_{ek}\hat R^{ck}{}_{fg}
+8H_{ab}{}^cH_d{}^{ef}\hat R^{ad}{}_{gh}\hat R^{bh}{}_{ek}\hat R^{kg}{}_{cf}
-4H_{abc}H^{def}\hat R^{ab}{}_{dg}\hat R^{ch}{}_{ek}\hat R^{gk}{}_{fh}
\nonumber\\
&
-4H_{abc}H^{def}\hat R^{ag}{}_{de}\hat R^{bh}{}_{fk}\hat R^{ck}{}_{gh}
-2H_{ab}{}^cH^{de}{}_f\hat R^{ab}{}_{gh}\hat R^{hk}{}_{de}\hat R^{fg}{}_{ck}
+H_{abc}H^{def}\hat R^{ab}{}_{de}\hat R^{cg}{}_{hk}\hat R^{hk}{}_{fg}
\nonumber\\
&
+H_{abc}H^{def}\hat R^{ab}{}_{gh}\hat R^{ck}{}_{de}\hat R^{gh}{}_{fk}
+H_{abc}H^{def}\hat R^{gh}{}_{de}\hat R^{ab}{}_{fk}\hat R^{ck}{}_{gh}
\Big)\,.
\nonumber
%
%
\end{align}
But we will work with a simpler form for these terms, which agrees with the above up to terms of order $H^4$, given in (\ref{eq:Lprime}). Finally, the remaining term cancels against the $\varepsilon_8$-term in (\ref{eq:e8e8H1}) provided that
\begin{equation}
c=\frac15\,,\quad d=\frac12\,,\quad e=-\frac12\,,
\end{equation}
fixing all remaining free coefficients in our ansatz (\ref{eq:F2R3}).

\subsection{Terms of order \texorpdfstring{$H^2$}{H**2}}
At this order we have from the $t_8$-terms (\ref{eq:t8-1}) and (\ref{eq:t8-2})\footnote{Here $(\nabla H)^{ab}{}_{cd}=\nabla^{[a}H^{b]}{}_{cd}$.}
\begin{align}
&{}
-12t_{a_1\cdots a_8}F^{a'}_{ab}\nabla^dF_{a'}^{a_1a_2}H^{ba_3a_4}(\nabla H)^{aca_5a_6}R^{a_7a_8}{}_{cd}
-12t_{a_1\cdots a_8}F^{a'}_{ab}\nabla^dF_{a'}^{a_1a_2}H^{ba_3a_4}R^{aca_5a_6}(\nabla H)_{cd}{}^{a_7a_8}
\nonumber\\
&{}
-6t_{a_1\cdots a_8}F^{a'}_{ab}\nabla^aF_{a'}^{a_1a_2}H^{ba_3a_4}(\nabla H)^{cda_5a_6}R^{a_7a_8}{}_{cd}
+(F^{a'}\leftrightarrow F_{a'})
\end{align}
and from the $\varepsilon_8$-terms (\ref{eq:e8-1})--(\ref{eq:e8-3})
\begin{equation}
-\frac{9}{4}\varepsilon_8\varepsilon_8(F_{a'}F^{a'})\nabla H\nabla HR
\sim
-\frac{9}{4}8!F_{a'}^{ab}\nabla^cF^{a'}_{[ab}H^d{}_{cd}\nabla^eH^f{}_{ef}R^{gh}{}_{gh]}\,.
%
%
\end{equation}
In addition we have the terms coming from the reduction of the $H^2R^3$-terms in (\ref{eq:Lprime}). It is not hard to see that these terms cannot cancel. It is therefore clear that one has to add terms of the form $H^2\nabla H^2R$ to the $D$-dimensional Lagrangian. However, if these terms have a contraction between the two $H$'s without derivatives they will give terms of the form $F^2\nabla H^2R$, but all the terms we need to cancel have the form $F\nabla F H\nabla HR$. Therefore, we should only add such terms if they can be integrated by parts to put one derivative on $F$. This shortens the list of possible terms. Taking a basis of such terms (see the appendix) one finds after a long calculation that to cancel all internal double Lorentz violating terms in the reduced action one should add to the $D$-dimensional Lagrangian the terms in (\ref{eq:Lbis}), where the terms involving a contraction of $HH$ or $\nabla H\nabla H$ are
\begin{equation}
\begin{aligned}
L_1=
&{}
-\frac32H^{abk}H_k{}^{gh}\nabla^{[c}H_{aef}\nabla^dH_{bgh}R^{ef]}{}_{cd}
-\frac{11}{4}H^{abg}H_{ef}{}^h\nabla^{[c}H_{ab}{}^{|k|}\nabla^dH_{kgh}R^{ef]}{}_{cd}
\\&{}
-3H^{abk}H_{keg}\nabla^{[c}H_{abh}\nabla^dH_f{}^{|gh|}R^{ef]}{}_{cd}
+5H_g{}^{ak}H_{ke}{}^b\nabla^{[c}H_{abh}\nabla^dH_f{}^{|gh|}R^{ef]}{}_{cd}
\\&{}
-\frac52H^{ab}{}_cH^{gh}{}_d\nabla^{[c}H_{ef}{}^{|k|}\nabla^dH_{kab}R^{ef]}{}_{gh}
+\frac{5}{12}H^{abk}H_k{}^{gh}\nabla^{[c}H_{def}\nabla^dH_{abc}R^{ef]}{}_{gh}
\\&{}
-5H_e{}^{ak}H_k{}^{gh}\nabla^{[c}H^{|b|}{}_{cd}\nabla^dH_{abf}R^{ef]}{}_{gh}
+2H^{abk}H_k{}^{gh}\nabla^{[c}H_{def}\nabla^dH_{abh}R^{ef]}{}_{cg}
\\&{}
-H^{abg}H_{ef}{}^h\nabla^{[c}H_{dh}{}^{|k|}\nabla^dH_{abk}R^{ef]}{}_{cg}
+11H^{abg}H_{ef}{}^h\nabla^{[c}H_{da}{}^{|k|}\nabla^dH_{bhk}R^{ef]}{}_{cg}
\\&{}
-2H_d{}^{ag}H^b{}_{ef}\nabla^{[c}H_{bhk}\nabla^dH_a{}^{|hk|}R^{ef]}{}_{cg}
-2H^{abk}H_k{}^{gh}\nabla^{[c}H_{bef}\nabla^dH_{adh}R^{ef]}{}_{cg}
\\&{}
-6H^{ab}{}_dH^{gh}{}_e\nabla^{[c}H_{fak}\nabla^dH_{bh}{}^{|k|}R^{ef]}{}_{cg}
-6H_d{}^{ak}H_k{}^{gb}\nabla^{[c}H_{aeh}\nabla^dH_{bf}{}^{|h|}R^{ef]}{}_{cg}
\\&{}
-H^{abk}H_{kge}\nabla^{[c}H_{acd}\nabla^dH_{bhf}R^{ef]gh}
-8H^{ab}{}_eH_{cdg}\nabla^{[c}H_{fak}\nabla^dH_{hb}{}^{|k|}R^{ef]gh}\,,
\end{aligned}
\label{eq:L1}
\end{equation}
while those containing no such contractions are
\begin{equation}
\begin{aligned}
L_2=
&{}
\frac52H_a{}^{bc}H^{def}\nabla^{[a}H_{bde}\nabla^kH_{cfk}R^{gh]}{}_{gh}
+6H^{abc}H^{de}{}_f\nabla^{[f}H_{cde}\nabla^kH_{abk}R^{gh]}{}_{gh}
\\&{}
+2H^{abc}H^{de}{}_f\nabla^{[f}H_{bdh}\nabla^gH_{cek}R^{hk]}{}_{ga}
-3H^{abc}H^{de}{}_f\nabla^{[f}H_{deh}\nabla^gH_{bck}R^{hk]}{}_{ga}
\\&{}
-11H^{abc}H^{de}{}_f\nabla^{[f}H_{bde}\nabla^gH_{chk}R^{hk]}{}_{ga}
-\frac23H^{ab}{}_cH^{def}\nabla^{[c}H_{def}\nabla^gH_{bhk}R^{hk]}{}_{ga}
\\&{}
-6H^{ab}{}_cH^{def}\nabla^{[c}H_{bde}\nabla^gH_{fhk}R^{hk]}{}_{ga}
-\frac52H^{ab}{}_cH^{de}{}_f\nabla^{[c}H_{bde}\nabla^fH_{ghk}R^{hk]g}{}_a
\\&{}
+\frac{11}{24}H^a{}_{bc}H^{def}\nabla^{[b}H_{def}\nabla^cH_{ghk}R^{hk]g}{}_a
+16H^{ab}{}_cH^{de}{}_f\nabla^{[c}H_{dhk}\nabla^fH_{beg}R^{hk]g}{}_a
\\&{}
+4H^a{}_{bc}H^{def}\nabla^{[b}H_{deh}\nabla^cH_{fgk}R^{hk]g}{}_a
+4H^{abc}H_{de}{}^f\nabla^{[d}H_{bch}\nabla^eH_{fgk}R^{hk]g}{}_a
\\&{}
-H^{abc}H^{de}{}_f\nabla^{[f}H_{dek}\nabla^gH_{cgh}R^{hk]}{}_{ab}
-3H^{abc}H^{de}{}_f\nabla^{[f}H_{cdk}\nabla^gH_{egh}R^{hk]}{}_{ab}
\\&{}
-8H_a{}^{bc}H^{def}\nabla^{[a}H_{efk}\nabla^gH_{bgh}R^{hk]}{}_{cd}
-6H_a{}^{bc}H^{de}{}_f\nabla^{[a}H_{bde}\nabla^fH_{cgk}R^{kh]g}{}_h
\\&{}
-\frac{11}{2}H_{ab}{}^cH^{def}\nabla^{[a}H_{kcf}\nabla^bH_{gde}R^{kh]g}{}_h
-12H_{ab}{}^cH^{def}\nabla^{[a}H_{kde}\nabla^bH_{gcf}R^{kh]g}{}_h
\\&{}
+3H_a{}^{bc}H^{de}{}_f\nabla^{[a}H_{kbc}\nabla^fH_{gde}R^{kh]g}{}_h
-8H_{ab}{}^cH^{de}{}_f\nabla^{[a}H_{gde}\nabla^bH_{chk}R^{fk]gh}
\\&{}
+32H_{ab}{}^cH^{de}{}_f\nabla^{[a}H_{gcd}\nabla^bH_{ehk}R^{fk]gh}
-\frac32H_{ab}{}^cH^{de}{}_f\nabla^{[a}H_{cde}\nabla^bH_{ghk}R^{fk]gh}
\\&{}
+\frac52H_{ab}{}^cH^{de}{}_f\nabla^{[a}H_{kcd}\nabla^bH_{egh}R^{fk]gh}
+\frac{1}{12}H_{abc}H^{def}\nabla^{[a}H_{kde}\nabla^bH_{fgh}R^{ck]gh}\,.
\end{aligned}
\label{eq:L2}
\end{equation}
Here we have written the answer as far as possible in terms of terms with an anti-symmetrization of four indices which allows them to be integrated by parts to put the reduced terms in the form $F\nabla FH\nabla HR$. We find only two terms left over which cannot be cast in this form, namely the first two terms in (\ref{eq:Lbis}). This result is highly non-unique due to the many ways one can integrate by parts and use Bianchi identities to rewrite it. Our strategy was to simplify the expressions for $L_1$ as much as possible first, before simplifying $L_2$, but the above expressions are probably not the best way to write these terms. Note that, ignoring factors of 2, the coefficients of the terms above involve only the prime factors 3, 5 or 11. This seems to suggest some substructure to these terms, but it is hard to say more without having a more systematic way to organize the terms.

We could now go on and consider terms of order $H^3$ in the reduced action. However, since we have already fixed all the possible terms in the $D$-dimensional action that are relevant for us these terms would have to cancel automatically. It would be nice to verify this as a consistency check, but we have not done so since the calculations are quite long, we have only checked that all terms in the reduced action which need to cancel can again be put in the form $F\nabla FH\nabla H^2$.

\section{Conclusions}
We have seen how to complete the $R^4$-terms in the tree-level string effective action by requiring that the effective action reduced to $D-d$ dimensions should have $O(d,d)$ symmetry.{ In fact, we showed that it is enough to require that all terms with an index contraction not compatible with $O(d,d)$, or rather an internal double Lorentz symmetry $O(d)\times O(d)$, cancel out.} We carried this out to fifth order in fields ignoring dilaton terms. It is in principle straightforward to extend this to compute all the couplings, though it requires some work. However, given the complicated structure of the $H^2\nabla H^2R$ terms, it would be important to first understand how to organize these terms. That the result is unique follows from \cite{Garousi:2020gio} and our result can be used as a guide to organizing the full (NS sector) completion of $R^4$ found there in a better way.

It is clear from our calculations that $O(d,d)$ symmetry appears due to very non-trivial cancellations in the reduced theory. Another important question is if it is possible to make the $O(d,d)$ symmetry more manifest already in $D$ dimension, probably at the expense of making Lorentz invariance less manifest.

\section*{Acknowledgments}
This work is supported by the grant ``Integrable Deformations'' (GA20-04800S) from the Czech Science Foundation (GA\v CR).

\newpage

\appendix
\section{Proof of (\ref{eq:eeR4})}\label{app:eeR4}
For completeness we give here a direct proof of the equality of the two expressions in (\ref{eq:eeR4}). We make the proof slightly more general by considering an arbitrary power of $R^{(-)}$.

\subsection{Vanishing of terms linear in the spin connection}
Let us first consider terms with one spin connection and show that they vanish up to total derivatives and higher order terms. They are
\begin{equation}
-n!\omega^{a_1a_2a_3}H_{[a_1a_2a_3}R^{(-)a_4a_5}{}_{a_4a_5}\cdots R^{(-)a_{n-1}a_n}{}_{a_{n-1}a_n]}
+(H\rightarrow-H)\,,
\end{equation}
which consists of a sum of terms with an even number of $H$'s of the form
\begin{equation}
\omega^{a_1a_2a_3}H_{[a_1a_2a_3}\nabla^{a_4}H^{a_5}{}_{a_4a_5}\cdots\nabla^{a_{6+2k}}H^{a_{7+2k}}{}_{a_{6+2k}a_{7+2k}}
R^{a_{8+2k}a_{9+2k}}{}_{a_{8+2k}a_{9+2k}}\cdots R^{a_{n-1}a_n}{}_{a_{n-1}a_n]}\,.
\end{equation}
This term can be rewritten as follows
\begin{equation}
\begin{aligned}
&\frac{n+1}{3}\eta^{a_5b}\omega^{a_1a_2a_3}H_{[a_1a_2a_3}\nabla^{a_4}H_{ba_4a_5}\left(\nabla H\cdots\nabla H\right)^{a_6\cdots a_{7+2k}}{}_{a_6\cdots a_{7+2k}}
\left(R\cdots R\right)^{a_{8+2k}\cdots a_n}{}_{a_{8+2k}\cdots a_n]}
\\
&{}
+\eta^{a_5b}\omega^{a_1a_2a_3}H_{b[a_1a_2}\nabla^{a_4}H_{a_3a_4a_5}
\left(\nabla H\cdots\nabla H\right)^{a_6\cdots a_{7+2k}}{}_{a_6\cdots a_{7+2k}}
\left(R\cdots R\right)^{a_{8+2k}\cdots a_n}{}_{a_{8+2k}\cdots a_n]}
\\
&{}
+\frac{4k}{3}\eta^{a_5b}\omega^{a_1a_2a_3}H_{[a_1a_2a_3}\nabla^{a_4}H_{a_4a_5a_6}\nabla^{a_6}H^{a_7}{}_{|b|a_7}
\left(\nabla H\cdots\nabla H\right)^{a_8\cdots a_{7+2k}}{}_{a_8\cdots a_{7+2k}}
\left(R\cdots R\right)^{a_{8+2k}\cdots a_n}{}_{a_{8+2k}\cdots a_n]}
\\
&{}
+\frac{n-5-4k}{3}\eta^{a_5b}\omega^{a_1a_2a_3}H_{[a_1a_2a_3}\nabla^{a_4}H_{a_{n-1}a_4a_5}
\left(\nabla H\cdots\nabla H\right)^{a_6\cdots a_{7+2k}}{}_{a_6\cdots a_{7+2k}}
\left(R\cdots R\right)^{a_{8+2k}\cdots a_n}{}_{a_{8+2k}\cdots|b|a_n]}\,.
\end{aligned}
\end{equation}
The first term is zero by the anti-symmetry in $a_5$ and $b$ and the last vanishes by the Bianchi identity for the last Riemann tensor. Integrating the second term by parts and dropping the total derivative and terms of higher order in fields it becomes minus the term we started with. This term is therefore given by $1/2$ times the third term, i.e.
\begin{equation}
\frac{2k}{9}\omega^{a_1a_2a_3}H_{[a_1a_2a_3}\nabla^{a_4}H_{a_4a_5a_6}\nabla_{a_7}H^{a_5a_6a_7}
\left(\nabla H\cdots\nabla H\right)^{a_8\cdots a_{7+2k}}{}_{a_8\cdots a_{7+2k}}
\left(R\cdots R\right)^{a_{8+2k}\cdots a_n}{}_{a_{8+2k}\cdots a_n]}\,.
\end{equation}
This vanishes if $k=0$. If $k>0$ we can apply the same trick as above to lower the $a_9$ index and following the same steps we find that the result vanishes unless $k>1$. Clearly, continuing in this way we find that the result must vanish to all orders in $H$. This completes the proof that the terms linear in the spin connection vanish modulo total derivatives and higher order terms, which are not relevant for our discussions here.

\subsection{Remaining terms}
Looking now at the terms quadratic in the spin connection, the first step is to note that
\begin{equation}
\begin{aligned}
&\varepsilon_{a_1\cdots a_n}\varepsilon^{b_1\cdots b_n}\omega^{(+)a_1a_2a_3}\omega^{(-)}_{b_1b_2b_3}R^{(-)a_4a_5}{}_{b_4b_5}\cdots R^{(-)a_{n-1}a_n}{}_{b_{n-1}b_n}
\\
=&\,{}
-n!
\omega^{(+)a_1a_2a_3}\omega^{(-)}_{[a_1a_2a_3}R^{(-)a_4a_5}{}_{a_4a_5}\cdots R^{(-)a_{n-1}a_n}{}_{a_{n-1}a_n]}
\\
=&\,{}
-(n+1)!\eta^{ba_1}\omega_{[b}^{(+)a_2a_3}\omega^{(-)}_{a_1a_2a_3}R^{(-)a_4a_5}{}_{a_4a_5}\cdots R^{(-)a_{n-1}a_n}{}_{a_{n-1}a_n]}
\\
&{}
-n!\eta^{ba_1}\omega_{[a_1}^{(+)a_2a_3}\omega^{(-)}_{|b|a_2a_3}R^{(-)a_4a_5}{}_{a_4a_5}\cdots R^{(-)a_{n-1}a_n}{}_{a_{n-1}a_n]}
\\
&{}
-2n!\eta^{ba_1}\omega_{[a_1}^{(+)a_2a_3}\omega^{(-)}_{a_2a_3|b|}R^{(-)a_4a_5}{}_{a_4a_5}\cdots R^{(-)a_{n-1}a_n}{}_{a_{n-1}a_n]}
\\
&{}
+(n-3)n!\eta^{ba_1}\omega_{[a_1}{}^{(+)a_2a_3}\omega^{(-)}_{a_2a_3a_4}R^{(-)a_4a_5}{}_{|b|a_5}\cdots R^{(-)a_{n-1}a_n}{}_{a_{n-1}a_n]}\,.
\end{aligned}
\end{equation}
The first term vanishes by the anti-symmetry in $a_1$ and $b$. Using the Bianchi identity $R^{(-)[abc]d}\sim-\frac13\partial^dH^{abc}$, where we neglected terms with more fields, and integrating by parts the last term becomes
\begin{equation}
\frac{n-3}{6}n!\omega^{(-)}_{[a_1a_2a_3}H^{a_1a_2a_3}R^{(-)a_4a_5}{}_{a_4a_5}\cdots R^{(-)a_{n-1}a_n}{}_{a_{n-1}a_n]}\,,
\end{equation}
where we dropped total derivative terms and terms of higher order in fields. Using this we have
\begin{equation}
\begin{aligned}
&\varepsilon_{a_1\cdots a_n}\varepsilon^{b_1\cdots b_n}
\left(\omega_{b_1}^{(+)a_1a_2}\omega^{(-)a_3}{}_{b_2b_3}-\omega^{(+)a_1a_2a_3}\omega^{(-)}_{b_1b_2b_3}\right)
R^{(-)a_4a_5}{}_{b_4b_5}\cdots R^{(-)a_{n-1}a_n}{}_{b_{n-1}b_n}
\\
\sim&\,{}
2n!\omega_{[a_1}^{(+)a_1a_2}\omega^{(-)}_{a_2a_3}{}^{a_3}\left(R^{(-)}\cdots R^{(-)}\right)^{a_4\cdots a_n}{}_{a_4\cdots a_n]}
-\frac{n-3}{6}n!\omega^{(-)}_{[a_1a_2a_3}H^{a_1a_2a_3}\left(R^{(-)}\cdots R^{(-)}\right)^{a_4\cdots a_n}{}_{a_4\cdots a_n]}
\end{aligned}
\end{equation}
and we may further rewrite the first term as
\begin{equation}
\begin{aligned}
&{}
-2n!\omega_{[a_1}^{(+)a_1a_2}H_{a_2a_3}{}^{a_3}\left(R^{(-)}\cdots R^{(-)}\right)^{a_4\cdots a_n}{}_{a_4\cdots a_n]}
\\&{}
+4(n-1)!\omega_{a_1}^{(+)[a_2|b|}\omega^{(+)}_{a_2b}{}^{a_1}\left(R^{(-)}\cdots R^{(-)}\right)^{a_4\cdots a_n]}{}_{a_4\cdots a_n}
\\&{}
-2(n-3)(n-1)!\omega_{a_1}^{(+)[a_1a_2}\omega^{(+)}_{a_2b}{}^{a_4}R^{(-)|b|a_5}{}_{a_4a_5}\left(R^{(-)}\cdots R^{(-)}\right)^{a_6\cdots a_n}{}_{a_6\cdots a_n]}
\\
\sim&\,{}
-2n!\omega^{(+)}_{[a_1}{}^{a_1a_2}H_{a_2a_3}{}^{a_3}\left(R^{(-)}\cdots R^{(-)}\right)^{a_4\cdots a_n}{}_{a_4\cdots a_n]}
\\&{}
+4(n-1)!\omega_{[a_1}^{(+)a_2b}\omega^{(+)}_{a_2|b|}{}^{a_1}\left(R^{(-)}\cdots R^{(-)}\right)^{a_4\cdots a_n}{}_{a_4\cdots a_n]}
\\&{}
-2(n-3)(n-1)!\omega_{[a_1}^{(+)a_1a_2}\partial_{a_4}\left[\omega^{(+)}_{a_2|b|}{}^{a_4}\omega_{a_5}^{(+)ba_5}\right]
\left(R^{(-)}\cdots R^{(-)}\right)^{a_6\cdots a_n}{}_{a_6\cdots a_n]}
\\
\sim&\,{}
-2n!\omega^{(+)}_{[a_1}{}^{a_1a_2}H_{a_2a_3}{}^{a_3}\left(R^{(-)}\cdots R^{(-)}\right)^{a_4\cdots a_n}{}_{a_4\cdots a_n]}
\\&{}
-(n+1)(n-1)!\omega_{[a_1}^{(+)a_1b}\omega^{(+)}_{a_2|b|}{}^{a_2}\left(R^{(-)}\cdots R^{(-)}\right)^{a_4\cdots a_n}{}_{a_4\cdots a_n]}\,.
\end{aligned}
\end{equation}
Adding now the total derivative term 
\begin{equation}
\begin{aligned}
&(n+1)(n-1)!\nabla^{(+)}_{[a_1}\left(\omega^{(+)}_{a_2}{}^{a_1a_2}R^{(-)a_3a_4}{}_{a_3a_4}\cdots R^{(-)a_{n-2}a_{n-1}}{}_{a_{n-2}a_{n-1}]}\right)
\\
=&\,{}
\frac{n+1}{2}(n-1)!\left(R^{(-)}\cdots R^{(-)}\right)^{a_1\cdots a_n}{}_{[a_1\cdots a_n]}
\\
&{}
+(n+1)(n-1)!\omega_{[a_1}^{(+)a_1b}\omega^{(+)}_{a_2|b|}{}^{a_2}\left(R^{(-)}\cdots R^{(-)}\right)^{a_4\cdots a_n}{}_{a_4\cdots a_n]}
\\
&{}
+\frac{n+1}{2}n!H_{[a_1a_2}{}^{a_1}\omega^{(+)}_{a_3}{}^{a_2a_3}\left(R^{(-)}\cdots R^{(-)}\right)^{a_4\cdots a_n}{}_{a_4\cdots a_n]}
\end{aligned}
\end{equation}
cancels the $\omega^2$-term and we are left with
\begin{equation}
\begin{aligned}
&\frac{n+1}{2}(n-1)!\left(R^{(-)}\cdots R^{(-)}\right)^{a_1\cdots a_n}{}_{[a_1\cdots a_n]}
+\frac{n-3}{2}n!\omega_{[a_1}^{(+)a_2a_3}H_{a_2a_3}{}^{a_1}\left(R^{(-)}\cdots R^{(-)}\right)^{a_4\cdots a_n}{}_{a_4\cdots a_n]}
\\
&{}
-\frac{n-3}{6}n!\omega^{(-)}_{[a_1a_2a_3}H^{a_1a_2a_3}\left(R^{(-)}\cdots R^{(-)}\right)^{a_4\cdots a_n}{}_{a_4\cdots a_n]}\,.
\end{aligned}
\end{equation}
Finally, we use the fact that
\begin{equation}
\begin{aligned}
&\omega^{(+)}_{[a_1}{}^{a_2a_3}H_{a_2a_3}{}^{a_1}\left(R^{(-)}\cdots R^{(-)}\right)^{a_4\cdots a_n}{}_{a_4\cdots a_n]}
\\
=&\,{}
\frac13\eta^{a_1b}\omega_b^{(+)a_2a_3}H_{[a_1a_2a_3}\left(R^{(-)}\cdots R^{(-)}\right)^{a_4\cdots a_n}{}_{a_4\cdots a_n]}
\\
&{}
+\frac{n-3}{3}\eta^{a_1b}\omega^{(+)}_{[a_1}{}^{a_2a_3}H_{a_4a_2a_3}R^{(-)a_4a_5}{}_{|b|a_5}\left(R^{(-)}\cdots R^{(-)}\right)^{a_6\cdots a_n}{}_{a_6\cdots a_n]}
\\
\sim&\,{}
\frac13\omega^{(+)a_1a_2a_3}H_{[a_1a_2a_3}\left(R^{(-)}\cdots R^{(-)}\right)^{a_4\cdots a_n}{}_{a_4\cdots a_n]}
\\
&{}
-\frac{n-3}{9}\omega^{(+)}_{[a_1}{}^{a_2a_3}H_{a_4a_2a_3}\partial_{a_5}H^{a_4a_5a_1}\left(R^{(-)}\cdots R^{(-)}\right)^{a_6\cdots a_n}{}_{a_6\cdots a_n]}
\\
\sim&\,{}
\frac13\omega^{(+)a_1a_2a_3}H_{[a_1a_2a_3}\left(R^{(-)}\cdots R^{(-)}\right)^{a_6\cdots a_n}{}_{a_6\cdots a_n]}
\\
&{}
+\frac{n-3}{18}H_{[a_1a_2a_3}H^{a_1a_2a_3}\left(R^{(-)}\cdots R^{(-)}\right)^{a_6\cdots a_n}{}_{a_6\cdots a_n]}
\end{aligned}
\label{eq:omega-H-id}
\end{equation}
and we get
\begin{equation}
\begin{aligned}
&
\frac{n+1}{2}(n-1)!\left(R^{(-)}\cdots R^{(-)}\right)^{a_1\cdots a_n}{}_{[a_1\cdots a_n]}
\\
&{}
+\frac{n-3}{6}n!\omega^{(+)a_1a_2a_3}H_{[a_1a_2a_3}\left(R^{(-)}\cdots R^{(-)}\right)^{a_4\cdots a_n}{}_{a_4\cdots a_n]}
\\
&{}
-\frac{n-3}{6}n!\omega^{(-)}_{[a_1a_2a_3}H^{a_1a_2a_3}\left(R^{(-)}\cdots R^{(-)}\right)^{a_4\cdots a_n}{}_{a_4\cdots a_n]}
\\
&{}
+\left(\frac{n-3}{6}\right)^2n!H_{[a_1a_2a_3}H^{a_1a_2a_3}\left(R^{(-)}\cdots R^{(-)}\right)^{a_4\cdots a_n}{}_{a_4\cdots a_n]}
\\
\sim&\,{}
\frac{n+1}{2}(n-1)!\left(R^{(-)}\cdots R^{(-)}\right)^{a_1\cdots a_n}{}_{[a_1\cdots a_n]}
\\
&{}
+\frac{n^2-9}{36}n!H^{a_1a_2a_3}H_{[a_1a_2a_3}\left(R^{(-)}\cdots R^{(-)}\right)^{a_4\cdots a_n}{}_{a_4\cdots a_n]}
\end{aligned}
\end{equation}
where we used our previous result which says that the terms linear in the spin connection vanish. Putting this together we have shown that
\begin{equation}
\begin{aligned}
&\tfrac{2}{n+1}\varepsilon_{a_1\cdots a_n}\varepsilon^{b_1\cdots b_n}
\left(\omega_{b_1}^{(+)a_1a_2}\omega^{(-)a_3}{}_{b_2b_3}-[\omega^{(+)}-\tfrac13H]^{a_1a_2a_3}[\omega^{(-)}+\tfrac13H]_{b_1b_2b_3}\right)
\left(R^{(-)}\cdots R^{(-)}\right)^{a_4\cdots a_n}{}_{b_4\cdots b_n}
\\
\sim&\,{}
(n-1)!\left(R^{(-)}\cdots R^{(-)}\right)^{a_1\cdots a_n}{}_{[a_1\cdots a_n]}
+\frac{n-1}{18}n!H^{a_1a_2a_3}H_{[a_1a_2a_3}\left(R^{(-)}\cdots R^{(-)}\right)^{a_4\cdots a_n}{}_{a_4\cdots a_n]}
\end{aligned}
\end{equation}
and setting $n=9$ we recover (\ref{eq:eeR4}).

\section{Details of cancellation of non-invariant terms}
Here we provide some further details of the calculations at order $H$ and $H^2$.

\subsection{Order \texorpdfstring{$H^1$}{H**1}}\label{app:B}
The $F^2\nabla HR^2$-terms in (\ref{eq:H1-1}) not involving any `traces' (contracted anti-symmetric pairs of indices) are rewritten as terms of the right structure, i.e. $F\nabla FHR^2$, plus terms involving at least one trace as follows
\begin{equation}
\begin{aligned}
t^{a_1\cdots a_8}F_{a'}^{ab}F^{a'}_{a_1a_2}\nabla^cH^d{}_{a_3a_4}&R_{daa_5a_6}R_{bca_7a_8}
\sim
8\nabla_h(F_{a'}^{ab}F^{a'ef})H_e{}^{cd}R_{dafg}R_{bc}{}^{gh}
\\
&{}
+2\nabla^f(F_{a'}^{ab}F^{a'}_{ef})H^{ecd}R_{dagh}R_{bc}{}^{gh}
-4F_{a'}^{ab}F^{a'}_{ef}\nabla^eH^{cgh}R_{bc}{}^{fd}R_{adgh}
\\
&{}
-4F_{a'}^{ab}F^{a'ef}\nabla^cH^{dgh}R_{bcfd}R_{aegh}
+4F_{a'}^{ab}F^{a'ef}\nabla^gH^{hcd}R_{bcef}R_{dagh}
\\
&{}
+8F_{a'}^{ab}F^{a'}_{ef}H^{ecd}R_{da}{}^{fg}\nabla^hR_{bcgh}
\end{aligned}
\end{equation}
and
\begin{equation}
\begin{aligned}
t^{a_1\cdots a_8}F_{a'}^{ab}F^{a'}_{a_1a_2}&\nabla_bH_{ca_3a_4}R^{cd}{}_{a_5a_6}R_{daa_7a_8}
\sim
\frac{5!}{2}F_{a'ab}F^{a'}_{ef}\nabla^{[b}H^{|che|}R^{fg}{}_{cd}R^{da]}{}_{gh}%
\\
&{}
+4F_{a'}^{ab}F^{a'ef}\nabla_eH^{cgh}R_{facd}R^d{}_{bgh}%
-4F_{a'}^{ab}F^{a'ef}\nabla_bH^{cgh}R_{dafc}R^d{}_{egh}%
\\
&{}
+4F_{a'}^{ab}F^{a'ef}\nabla^cH_{egh}R_{facd}R_b{}^{dgh}%
+2F_{a'}^{ab}F^{a'ef}\nabla_bH_{egh}R_{afcd}R^{cdgh}%
\\
&{}
-4F_{a'}^{ab}F^{a'ef}\nabla_gH_{hce}R_{ab}{}^{cd}R_{fd}{}^{gh}%
+4F_{a'}^{ab}F^{a'ef}\nabla_cH_{hef}R_{ag}{}^{cd}R_{bd}{}^{gh}%
\\
&{}
-2F_{a'}^{ab}F^{a'ef}\nabla_bH_{cef}R^{cdgh}R_{dagh}%
-2F_{a'}^{ab}F^{a'ef}\nabla_bH_{cgh}R^{cd}{}_{ef}R_{da}{}^{gh}%
\\
&{}
-2F_{a'}^{ab}F^{a'ef}\nabla_bH_{cgh}R^{cdgh}R_{daef}%
+4F_{a'}^{ab}F^{a'ef}\nabla_bH_{che}R_{afg}{}^hR^{gc}%
\\
&{}
+2\cdot3!F_{a'ab}F^{a'}_{ef}\nabla^bH^{che}R^{[fd}{}_{cd}R^{a]}{}_h%
+2\cdot3!F_{a'ab}F^{a'}_{ef}\nabla^dH^{che}R^{[ab}{}_{cd}R^{f]}{}_h%
\\
&{}
-3!F_{a'ab}F^{a'}_{ef}\nabla^fH^{che}R^{[ab}{}_{cd}R^{d]}{}_h%
-2F_{a'ab}F^{a'}_{ef}\nabla^fH^{che}R^{ab}{}_{gh}R^g{}_c\,.
\end{aligned}
\end{equation}
The last five terms are proportional to the equations of motion, modulo terms we are ignoring, and can be dropped. Continuing in this way one eventually finds that the $t_8$-terms can be written as in (\ref{eq:H1-t8-simpl}) with $Y$ given by
\begin{equation}
\begin{aligned}
Y^{a'abdef}
=\,&{}
8\nabla_cF^{a'dh}R^a{}_{hg}{}^eR^{bgfc}
-16\nabla_cF^{a'dh}R^a{}_{hg}{}^eR^{bfcg}
-8\nabla_cF^{a'dh}R^{ae}{}_{hg}R^{bfcg}
\\
&{}
-16\nabla_cF^{a'dh}R^{ae}{}_{hg}R^{bgfc}
+8\nabla^aF^{a'dh}R^{bcge}R_{hcg}{}^f
-4\nabla_cF^{a'}_{gh}R^{agcd}R^{bhef}
\\
&{}
-4\nabla_cF^{a'dh}R^{ec}{}_{hg}R^{abfg}
+4\nabla^cF^{a'dh}R^a{}_{hgc}R^{bgef}
-4\nabla^cF^{a'de}R^a{}_{ghc}R^{bghf}
\\
&{}
-4\nabla^aF^{a'cd}R^{begh}R^f{}_{cgh}
+4\nabla^aF^{a'}_{gh}R^{dgbc}R^{efh}{}_c
+2\nabla_cF^{a'ef}R^{acgh}R^{bd}{}_{gh}
\\
&{}
-2\nabla_cF^{a'dg}R^{chef}R^{ab}{}_{gh}
-2\nabla^aF^{a'ef}R^{bcgh}R^d{}_{cgh}
+\nabla_cF^{a'}_{gh}R^{abde}R^{ghcf}
\\
&{}
-2\nabla^aF^{a'}_{gh}R^{ghcd}R_c{}^{bef}
+\nabla_cF^{a'ef}R^{abgh}R^{cd}{}_{gh}
-2\nabla^aF^{a'dc}R^b{}_{cgh}R^{efgh}
\\
&{}
+\nabla_cF^{a'}_{gh}R^{abcd}R^{efgh}
\end{aligned}
\label{eq:Y}
\end{equation}
and similarly for $Y_{a'}^{abdef}$ with the primed index lowered.

\subsection{Order \texorpdfstring{$H^2$}{H**2}}
We use the following basis for the $H^2\nabla H^2R$ terms involving a contraction of $H$ with $H$ or $\nabla H$ with $\nabla H$ (contractions with the index on the derivative do not count)\footnote{The index placement here is chosen purely for readability.}
\begin{equation}
\begin{aligned}
&f_1=4!H_{dab}H_{cgh}\nabla^{[c}H_{kef}\nabla^dH_{kab}R^{ef]}{}_{gh}\\
&f_2=4!H_{kef}H_{cgh}\nabla^{[c}H_{dab}\nabla^dH_{kab}R^{ef]}{}_{gh}\\
&f_3=4!H_{kab}H_{kgh}\nabla^{[c}H_{def}\nabla^dH_{abc}R^{ef]}{}_{gh}\\
&f_4=4!H_{kab}H_{kgh}\nabla^{[c}H_{acd}\nabla^dH_{bef}R^{ef]}{}_{gh}\\
&f_5=4!H_{kcd}H_{kgh}\nabla^{[c}H_{abe}\nabla^dH_{abf}R^{ef]}{}_{gh}\\
&f_6=4!H_{kea}H_{kgh}\nabla^{[c}H_{bcd}\nabla^dH_{abf}R^{ef]}{}_{gh}\\
&f_7=4!H_{def}H_{kgh}\nabla^{[c}H_{kab}\nabla^dH_{abc}R^{ef]}{}_{gh}\\
&f_8=4!H_{acd}H_{kgh}\nabla^{[c}H_{kab}\nabla^dH_{bef}R^{ef]}{}_{gh}\\
&f_9=4!H_{acd}H_{kgh}\nabla^{[c}H_{bke}\nabla^dH_{abf}R^{ef]}{}_{gh}\\
&f_{10}=4!H_{def}H_{ghk}\nabla^{[c}H_{abh}\nabla^dH_{abk}R^{ef]}{}_{cg}\\
&f_{11}=4!H_{abh}H_{ghk}\nabla^{[c}H_{def}\nabla^dH_{abk}R^{ef]}{}_{cg}\\
&f_{12}=4!H_{efh}H_{gab}\nabla^{[c}H_{dhk}\nabla^dH_{abk}R^{ef]}{}_{cg}\\
&f_{13}=4!H_{efh}H_{kgh}\nabla^{[c}H_{dab}\nabla^dH_{abk}R^{ef]}{}_{cg}\\
&f_{14}=4!H_{efh}H_{gbk}\nabla^{[c}H_{dab}\nabla^dH_{hka}R^{ef]}{}_{cg}\\
&f_{15}=4!H_{efh}H_{kgd}\nabla^{[c}H_{hab}\nabla^dH_{kab}R^{ef]}{}_{cg}\\
&f_{16}=4!H_{dab}H_{gbk}\nabla^{[c}H_{hef}\nabla^dH_{hka}R^{ef]}{}_{cg}\\
&f_{17}=4!H_{kab}H_{kgd}\nabla^{[c}H_{hef}\nabla^dH_{hab}R^{ef]}{}_{cg}\\
&f_{18}=4!H_{hab}H_{gbk}\nabla^{[c}H_{hef}\nabla^dH_{dka}R^{ef]}{}_{cg}\\
&f_{19}=4!H_{kab}H_{kgh}\nabla^{[c}H_{hef}\nabla^dH_{abd}R^{ef]}{}_{cg}\\
&f_{20}=4!H_{kab}H_{gab}\nabla^{[c}H_{hef}\nabla^dH_{kdh}R^{ef]}{}_{cg}\\
&f_{21}=4!H_{dab}H_{gef}\nabla^{[c}H_{ahk}\nabla^dH_{bhk}R^{ef]}{}_{cg}\\
&f_{22}=4!H_{hab}H_{gef}\nabla^{[c}H_{dhk}\nabla^dH_{kab}R^{ef]}{}_{cg}\\
&f_{23}=4!H_{dhk}H_{ghf}\nabla^{[c}H_{abe}\nabla^dH_{kab}R^{ef]}{}_{cg}\\
&f_{24}=4!H_{dab}H_{ghf}\nabla^{[c}H_{keh}\nabla^dH_{kab}R^{ef]}{}_{cg}
\end{aligned}
\qquad
\begin{aligned}
&f_{25}=4!H_{dab}H_{ghf}\nabla^{[c}H_{ebk}\nabla^dH_{hka}R^{ef]}{}_{cg}\\
&f_{26}=4!H_{akd}H_{kgh}\nabla^{[c}H_{abe}\nabla^dH_{bfh}R^{ef]}{}_{cg}\\
&f_{27}=4!H_{kdh}H_{kgh}\nabla^{[c}H_{abe}\nabla^dH_{abf}R^{ef]}{}_{cg}\\
&f_{28}=4!H_{def}H_{gbk}\nabla^{[c}H_{abc}\nabla^dH_{hka}R^{ef]}{}_{gh}\\
&f_{29}=4!H_{def}H_{kgc}\nabla^{[c}H_{kab}\nabla^dH_{abh}R^{ef]}{}_{gh}\\
&f_{30}=4!H_{abc}H_{gbk}\nabla^{[c}H_{def}\nabla^dH_{hka}R^{ef]}{}_{gh}\\
&f_{31}=4!H_{kab}H_{kgc}\nabla^{[c}H_{def}\nabla^dH_{abh}R^{ef]}{}_{gh}\\
&f_{32}=4!H_{kcd}H_{bkg}\nabla^{[c}H_{aef}\nabla^dH_{abh}R^{ef]}{}_{gh}\\
&f_{33}=4!H_{kcd}H_{gef}\nabla^{[c}H_{kab}\nabla^dH_{abh}R^{ef]}{}_{gh}\\
&f_{34}=4!H_{kcd}H_{bgf}\nabla^{[c}H_{ake}\nabla^dH_{abh}R^{ef]}{}_{gh}\\
&f_{35}=4!H_{kcd}H_{kgf}\nabla^{[c}H_{abe}\nabla^dH_{abh}R^{ef]}{}_{gh}\\
&f_{36}=4!H_{kcd}H_{agf}\nabla^{[c}H_{abe}\nabla^dH_{bkh}R^{ef]}{}_{gh}\\
&f_{37}=4!H_{kcd}H_{age}\nabla^{[c}H_{kab}\nabla^dH_{bhf}R^{ef]}{}_{gh}\\
&f_{38}=4!H_{eab}H_{gfa}\nabla^{[c}H_{kcd}\nabla^dH_{hkb}R^{ef]}{}_{gh}\\
&f_{39}=4!H_{kab}H_{age}\nabla^{[c}H_{kcd}\nabla^dH_{bhf}R^{ef]}{}_{gh}\\
&f_{40}=4!H_{eab}H_{gcd}\nabla^{[c}H_{fbk}\nabla^dH_{hka}R^{ef]}{}_{gh}\\
&f_{41}=4!H_{abc}H_{ged}\nabla^{[c}H_{abk}\nabla^dH_{khf}R^{ef]}{}_{gh}\\
&f_{42}=4!H_{abc}H_{bge}\nabla^{[c}H_{adk}\nabla^dH_{khf}R^{ef]}{}_{gh}\\
&f_{43}=4!H_{kab}H_{kgh}\nabla^{[c}H_{aef}\nabla^dH_{bef}R^{gh]}{}_{cd}\\
&f_{44}=4!H_{kab}H_{fgh}\nabla^{[c}H_{abe}\nabla^dH_{efk}R^{gh]}{}_{cd}\\
&f_{45}=4!H_{kab}H_{geb}\nabla^{[c}H_{kaf}\nabla^dH_{hef}R^{gh]}{}_{cd}\\
&f_{46}=4!H_{kab}H_{gab}\nabla^{[c}H_{kef}\nabla^dH_{hef}R^{gh]}{}_{cd}\\
&f_{47}=4!H^{kae}H_{gab}\nabla^{[c}H_{kbf}\nabla^dH^{hef}R^{gh]}{}_{cd}\\
&{}
\end{aligned}
\end{equation}
and those without such contractions
\begin{equation}
\begin{aligned}
&g_1=4!H_{abc}H_{def}\nabla^{[a}H_{bef}\nabla^kH_{cdk}R^{gh]}{}_{gh}\\
&g_2=4!H_{abc}H_{def}\nabla^{[a}H_{def}\nabla^kH_{bck}R^{gh]}{}_{gh}\\
&g_3=4!H_{abc}H_{def}\nabla^{[d}H_{aef}\nabla^kH_{bck}R^{gh]}{}_{gh}\\
&g_4=4!H_{abc}H_{def}\nabla^{[e}H_{bdh}\nabla^gH_{cfk}R^{hk]}{}_{ag}\\
&g_5=4!H_{abc}H_{def}\nabla^{[e}H_{bcd}\nabla^gH_{fhk}R^{hk]}{}_{ag}\\
&g_6=4!H_{abc}H_{def}\nabla^{[d}H_{efh}\nabla^gH_{bck}R^{hk]}{}_{ag}\\
&g_7=4!H_{abc}H_{def}\nabla^{[b}H_{efh}\nabla^gH_{cdk}R^{hk]}{}_{ag}\\
&g_8=4!H_{abc}H_{def}\nabla^{[d}H_{bef}\nabla^gH_{chk}R^{hk]}{}_{ag}\\
&g_9=4!H_{abc}H_{def}\nabla^{[b}H_{def}\nabla^gH_{chk}R^{hk]}{}_{ag}\\
&g_{10}=4!H_{abc}H_{def}\nabla^{[b}H_{cef}\nabla^gH_{dhk}R^{hk]}{}_{ag}\\
&g_{11}=4!H_{abc}H_{def}\nabla^{[e}H_{bcd}\nabla^fH_{ghk}R^{hk]}{}_{ag}\\
&g_{12}=4!H_{abc}H_{def}\nabla^{[b}H_{cef}\nabla^dH_{ghk}R^{hk]}{}_{ag}\\
&g_{13}=4!H_{abc}H_{def}\nabla^{[b}H_{def}\nabla^cH_{ghk}R^{hk]}{}_{ag}\\
&g_{14}=4!H_{abc}H_{def}\nabla^{[b}H_{chk}\nabla^dH_{efg}R^{hk]}{}_{ag}\\
&g_{15}=4!H_{abc}H_{def}\nabla^{[b}H_{dhk}\nabla^cH_{efg}R^{hk]}{}_{ag}\\
&g_{16}=4!H_{abc}H_{def}\nabla^{[b}H_{dhk}\nabla^eH_{cfg}R^{hk]}{}_{ag}\\
&g_{17}=4!H_{abc}H_{def}\nabla^{[e}H_{bhk}\nabla^fH_{cdg}R^{hk]}{}_{ag}\\
&g_{18}=4!H_{abc}H_{def}\nabla^{[e}H_{dhk}\nabla^fH_{bcg}R^{hk]}{}_{ag}\\
&g_{19}=4!H_{abc}H_{def}\nabla^{[b}H_{efh}\nabla^cH_{dgk}R^{hk]}{}_{ag}\\
&g_{20}=4!H_{abc}H_{def}\nabla^{[b}H_{efh}\nabla^dH_{cgk}R^{hk]}{}_{ag}\\
&g_{21}=4!H_{abc}H_{def}\nabla^{[b}H_{cdh}\nabla^eH_{fgk}R^{hk]}{}_{ag}\\
&g_{22}=4!H_{abc}H_{def}\nabla^{[e}H_{bdh}\nabla^fH_{cgk}R^{hk]}{}_{ag}
\end{aligned}
\qquad
\begin{aligned}
&g_{23}=4!H_{abc}H_{def}\nabla^{[d}H_{bch}\nabla^eH_{fgk}R^{hk]}{}_{ag}\\
&g_{24}=4!H_{abc}H_{def}\nabla^{[g}H_{efk}\nabla^dH_{cgh}R^{hk]}{}_{ab}\\
&g_{25}=4!H_{abc}H_{def}\nabla^{[c}H_{efk}\nabla^gH_{dgh}R^{hk]}{}_{ab}\\
&g_{26}=4!H_{abc}H_{def}\nabla^{[e}H_{cdk}\nabla^gH_{fgh}R^{hk]}{}_{ab}\\
&g_{27}=4!H_{abc}H_{def}\nabla^{[b}H_{efk}\nabla^gH_{cgh}R^{hk]}{}_{ad}\\
&g_{28}=4!H_{abc}H_{def}\nabla^{[b}H_{cek}\nabla^gH_{fgh}R^{hk]}{}_{ad}\\
&g_{29}=4!H_{abc}H_{def}\nabla^{[d}H_{abc}\nabla^eH_{fgk}R^{kh]}{}_{gh}\\
&g_{30}=4!H_{abc}H_{def}\nabla^{[a}H_{bef}\nabla^cH_{dgk}R^{kh]}{}_{gh}\\
&g_{31}=4!H_{abc}H_{def}\nabla^{[a}H_{bef}\nabla^dH_{cgk}R^{kh]}{}_{gh}\\
&g_{32}=4!H_{abc}H_{def}\nabla^{[a}H_{kcd}\nabla^bH_{gef}R^{kh]}{}_{gh}\\
&g_{33}=4!H_{abc}H_{def}\nabla^{[a}H_{kef}\nabla^bH_{gcd}R^{kh]}{}_{gh}\\
&g_{34}=4!H_{abc}H_{def}\nabla^{[a}H_{kbc}\nabla^dH_{gef}R^{kh]}{}_{gh}\\
&g_{35}=4!H_{abc}H_{def}\nabla^{[a}H_{kbe}\nabla^dH_{gcf}R^{kh]}{}_{gh}\\
&g_{36}=4!H_{abc}H_{def}\nabla^{[a}H_{gef}\nabla^bH_{dhk}R^{ck]}{}_{gh}\\
&g_{37}=4!H_{abc}H_{def}\nabla^{[a}H_{gef}\nabla^bH_{chk}R^{dk]}{}_{gh}\\
&g_{38}=4!H_{abc}H_{def}\nabla^{[a}H_{gce}\nabla^bH_{fhk}R^{dk]}{}_{gh}\\
&g_{39}=4!H_{abc}H_{def}\nabla^{[a}H_{def}\nabla^bH_{ghk}R^{ck]}{}_{gh}\\
&g_{40}=4!H_{abc}H_{def}\nabla^{[a}H_{cef}\nabla^bH_{ghk}R^{dk]}{}_{gh}\\
&g_{41}=4!H_{abc}H_{def}\nabla^{[a}H_{kef}\nabla^bH_{dgh}R^{ck]}{}_{gh}\\
&g_{42}=4!H_{abc}H_{def}\nabla^{[a}H_{kce}\nabla^bH_{fgh}R^{dk]}{}_{gh}\\
&g_{43}=4!H_{abc}H_{def}\nabla^{[a}H_{kef}\nabla^bH_{cgh}R^{dk]}{}_{gh}\\
&{}
\end{aligned}
\end{equation}
Adding a linear combination of these
\begin{equation}
\sum_i c_if_i
+\sum_i d_ig_i\,,
\end{equation}
to the $D$-dimensional Lagrangian one finds after a long calculation that all the terms violating the internal double Lorentz symmetry cancel if we take the following non-zero coefficients
\begin{equation}
\begin{aligned}
&c_1=\frac{15}{4}\,,\qquad
c_3=\frac58\,,\qquad
c_6=-\frac{15}{2}\,,\qquad
c_{11}=-3\,,\qquad
c_{12}=-\frac32\,,\qquad
c_{14}=\frac{33}{2}\,,
\\
&c_{15}=-3\,,\qquad
c_{18}=-3\,,\qquad
c_{25}=-9\,,\qquad
c_{26}=-9\,,\qquad
c_{39}=\frac32\,,\qquad
c_{40}=-12\,,
\\
&c_{43}=-\frac94\,,\qquad
c_{44}=\frac{51}{8}\,,\qquad
c_{45}=-\frac92\,,\qquad
c_{47}=-\frac{15}{2}
\end{aligned}
\end{equation}
and
\begin{equation}
\begin{aligned}
&d_1=\frac{15}{4}\,,\qquad
d_3=9\,,\qquad
d_4=3\,,\qquad
d_6=\frac92\,,\qquad
d_8=\frac{33}{2}\,,\qquad
d_9=-1\,,
\\
&d_{10}=-9\,,\qquad
d_{12}=-\frac{15}{4}\,,\qquad
d_{13}=-\frac{11}{16}\,,\qquad
d_{16}=-24\,,\qquad
d_{19}=-6\,,\qquad
d_{23}=-6\,,
\\
&
d_{24}=\frac32\,,\qquad
d_{26}=\frac92\,,\qquad
d_{27}=-12\,,\qquad
d_{31}=-9\,,\qquad
d_{32}=-\frac{33}{4}\,,\qquad
d_{33}=-18\,,
\\
&
d_{34}=\frac92\,,\qquad
d_{37}=-12\,,\qquad
d_{38}=48\,,\qquad
d_{40}=-\frac94\,,\qquad
d_{41}=\frac18\,,\qquad
d_{42}=\frac{15}{4}
\end{aligned}
\end{equation}
and add the two terms in (\ref{eq:Lbis}) without the anti-symmetization in the indices. Here we have tried to pick a minimal solution by first setting as many of the $c_i$'s as possible to zero, though there may exist a better choice of solution. The solution then takes the form of (\ref{eq:Lbis}).

\bibliographystyle{nb}
\bibliography{biblio}{}
\end{document}